\documentclass[12pt,preprint]{aastex}

\newcommand{\tlusty}{{\sc Tlusty}}

\newcommand{\teff}{$T_{\rm eff}$}
\newcommand{\grid}{{\sc Bstar2006}}

\newcommand{\kms}{km\,s$^{-1}$}

\newcommand{\vtur}{$V_{\rm t}$}

\shorttitle{NLTE Model Atmospheres of B-type Stars.}
\shortauthors{Lanz \& Hubeny}

\begin{document}

\title{A Grid of NLTE Line-Blanketed Model Atmospheres \\
       of Early B-type Stars}

\author{Thierry Lanz}
\affil{Department of Astronomy, University of Maryland, College Park, MD 20742;
   lanz@astro.umd.edu}

\and

\author{Ivan Hubeny}
\affil{Steward Observatory, University of Arizona, Tucson, AZ 85721;
    hubeny@as.arizona.edu}


\begin{abstract}
We have constructed a comprehensive grid of 1540 metal 
line-blanketed, NLTE, plane-parallel, hydrostatic model
atmospheres for the basic parameters appropriate to early B-type stars.
The \grid\ grid considers 16 values of effective temperatures, 
$15\,000$\,K $\leq$ \teff\ $\leq\,30\,000$\,K with 1\,000\,K steps, 
13 surface gravities, $1.75\leq\log g\leq 4.75$ with  
0.25\,dex steps, 6 chemical compositions, and a microturbulent velocity
of 2\,\kms.  The lower limit of $\log g$ for a given effective temperature
is set by an approximate location of the Eddington limit.
The selected chemical compositions range from twice to one tenth
of the solar metallicity and metal-free. Additional
model atmospheres for B supergiants ($\log g\leq 3.0$) have been
calculated with a higher microturbulent velocity (10\,\kms) and
a surface composition that is enriched in helium and nitrogen,
and depleted in carbon.
This new grid complements our earlier {\sc Ostar2002} grid of O-type stars
(Lanz~\& Hubeny, 2003, ApJS, 146, 417). The paper contains a description
of the \grid\ grid and some illustrative examples and comparisons. 
NLTE ionization fractions, bolometric corrections, radiative accelerations,
and effective gravities are obtained over the parameter range covered by the grid.
By extrapolating radiative accelerations, we have determined an improved
estimate of the Eddington limit in absence of rotation between 55\,000 and
15\,000\,K. The complete \grid\ grid is available at the \tlusty\ website.
\end{abstract}



\section{INTRODUCTION}
\label{IntroSect}

This paper is the second of a series dealing with new grids of non-LTE (NLTE),
metal line-blanketed model atmospheres of hot stars. Paper~I was devoted
to O-type stars \citep[][{\sc Ostar2002}]{OS02}, covering a range of effective
temperatures from 27\,500 to 55\,000\,K and 10 chemical compositions from
metal-rich relative to the Sun to metal-free. The new grid presented in this
paper complements this initial grid by extending it to cooler temperatures
($15\,000\leq$~\teff~$\leq 30\,000$\,K). Together, these two grids provide
a full set of model atmospheres necessary to construct composite model spectra
of young clusters of massive stars, OB associations, and starburst galaxies.
We plan a third forthcoming paper to further extend these two grids of NLTE
line-blanketed model atmospheres to late B-type and early A-type stars.

Advances in numerical schemes applied to stellar atmosphere modeling and large
increases in computational resources during the last decade have provided the
ability to calculate essentially ``exact'', fully blanketed NLTE hydrostatic
model stellar atmospheres. Earlier model atmospheres of B-type stars neglected
either departures from Local Thermodynamic Equilibrium (LTE; Kurucz 1993) or
the effect of line opacity from heavier species such as the initial H-He NLTE
model atmospheres \citep{mihalas70}.
Major improvements are thus expected from the inclusion of a detailed treatment
of line opacity without assuming LTE since these two ingredients are important
in hot stellar atmospheres. These advances have been well documented with
applications of the {\sc Ostar2002} grid to the analysis of O stars in the
Galaxy and the Small Magellanic Cloud (SMC; e.~g., Bouret et~al. 2003,
Heap et~al. 2006, Lanz et~al. 2007). The new model atmospheres predict and
consistently match lines of different ions and species, hence robustly supporting
the newly derived properties of O stars such as effective temperatures that are
systematically lower than those derived using unblanketed model atmospheres.
Furthermore, because the new model atmospheres incorporate the contribution
of all significant species to the total opacity, NLTE abundance studies will
include all background opacities in contrast to earlier line-formation studies
based on NLTE model atmospheres that incorporated only hydrogen and helium
besides the studied species. These new NLTE model atmospheres should 
therefore predict ionization balances with significantly better precision,
hence enhancing the accuracy of abundance determinations from lines of several ions.

In Paper~I, we discussed extensively the reasons why hydrostatic model
atmospheres are relevant for O star studies despite neglecting the stellar
wind. We argued in particular that the basic properties of massive stars
are determined on much safer grounds using selected lines formed
in the quasi-static photosphere rather than from lines formed in the
supersonic wind. Indeed, many uncertainties still remain in current wind models
such as wind clumping \citep{bouret05} and wind ionization by X-rays
(e.~g., Martins et~al. 2005). {\sc Ostar2002} models are frequently adopted
now to describe the density structure of the inner, quasi-static
layers of unified models of O stars, such as the CMFGEN models \citep{CMFGEN}.

On the main sequence, the radiatively-driven stellar winds observed in O-type stars
are vanishing very quickly in early B-type stars (around B0-B1 spectral type;
that is, \teff~$\la$~30\,000\,K).
Using hydrostatic model atmospheres to study B dwarfs is therefore fully
appropriate. The case of B supergiants may however differ because
of their large spherical extension and wind. \cite{dufton05} analyzed several
luminous B supergiants in the SMC. They derived similar results from the
hydrostatic TLUSTY model atmospheres and from the unified FASTWIND wind models
\citep{fastwind97,herrero02}. Their careful discussion thus provides adequate grounds for
using TLUSTY models in these spectral analyses, keeping within the same limits
as those set for O-type stars.

The paper is organized as follows. Section~\ref{ModelSect} recalls the basic assumptions
made in TLUSTY, the numerical methods used, and recent changes implemented in
the program. The bulk of atomic data used in the model calculations are the
same as those adopted for the {\sc Ostar2002} models, and \S\ref{AtDataSect}
is restricted to only document the adopted changes or improvements in model atoms. 
We describe briefly the grid in \S\ref{GridSect}
and we present some illustrative results in \S\ref{ResSect}.


\section{TLUSTY MODEL ATMOSPHERES}
\label{ModelSect}

The computer program TLUSTY \citep{tlusty88, NLTE1} has been developed for
calculating model stellar atmospheres assuming an atmosphere with a plane-parallel
geometry, in hydrostatic equilibrium, in radiative equilibrium
(or in radiative-convective equilibrium for cool stellar atmospheres),
and allowing for departures from LTE for an arbitrary set of chemical species
represented by model atoms of various complexity.
TLUSTY has been designed so that it allows for a total flexibility in selecting
and treating chemical species and opacities. Because of limited
computational resources, the original idea was to incorporate explicitly in 
the NLTE calculation only
the most important opacity sources (in hot stellar atmospheres, H and He mainly;
C, N, and O, in later stages). However, the introduction of the hybrid Complete
Linearization/Accelerated Lambda Iteration \citep{NLTE1} method has enabled us
to build model atmospheres accounting for all significant species and opacities
(continua as well as lines) without assuming LTE --- the so-called NLTE
fully-blanketed model atmospheres. \cite{OS02} have presented the first extensive
grid of such model atmospheres for the range of stellar parameters covered
by O-type stars. In Paper~I, we have described the physical assumptions and numerical
methods used to calculate these model atmospheres. In the present study, we
use the same methods and, in particular, we use opacity sampling (OS) to represent
the complicated frequency-dependent iron line opacity. Because of the smaller thermal
Doppler width and the lower microturbulent velocity (2\,\kms), sampling the spectrum
with the same characteristic frequency step (0.75 fiducial Doppler width) thus
requires many more frequencies over the whole spectrum than in the {\sc Ostar2002}
models --- typically, the \grid\  models consider about 380\,000 frequencies.
For more details on the methods, the reader is referred to Paper~I. 

The {\sc Ostar2002} models have been calculated with TLUSTY, version 198, while
we have used a later version, v. 201, for the \grid\  grid. The newer version of 
TLUSTY is a unified program that can calculate accretion disk models as well as
model stellar atmospheres (two separate codes existed for v.~198). Other additions
consist in a treatment of Compton scattering, opacities at high energies, improvements
for convection in cool models, and minor changes and fixes. These changes are
however not directly relevant to the modeling of O and B-type stars.


\section{ATOMIC DATA}
\label{AtDataSect}

We have closely followed the approach adopted in \cite{OS02} regarding the treatment
and inclusion of atomic data in the model atmosphere calculations. In particular,
the bulk of the data are taken from Topbase, the Opacity Project database
\citep{IOP95, IOP97}.\footnote{{\tt http://vizier.u-strasbg.fr/topbase/}} The level
energies have been systematically updated with the more accurate experimental energies extracted
from the Atomic and Spectroscopic Database
at NIST.\footnote{{\tt http://physics.nist.gov/PhysRefData/ASD/index.html}}
Finally, we use the extensive Kurucz iron line
data \citep{CD22}.\footnote{{\tt http://kurucz.harvard.edu/cdroms.html}}
For a detailed description, refer to \S 4 of Paper~I.

Compared to the {\sc Ostar2002} grid, we consider lower ions explicitly
in the NLTE calculations because we anticipate lower ionization from the lower
effective temperatures. For instance, detailed models of some neutral atoms
(\ion{C}{1}, \ion{N}{1}, \ion{O}{1}, \ion{Ne}{1}) are included in \grid\  models,
while some of the highest ions such as \ion{O}{6} are skipped. The
ionization balance of the different species (see \S\ref{ResSect}) provides
a good justification of this choice. We have used the same model atoms for most
ions in common with the {\sc Ostar2002} grid. In some instances, we have however
expanded the model atoms to include all levels below the ionization limit as they are
listed in OP (hence, completion up to $n=10$ as in OP).
The following subsections provide the relevant details of new or updated model atoms ---
refer to Paper~I for model atoms not discussed here. 

Table~\ref{IonTbl} summarizes the atomic data included in the model atmospheres,
as well as the references to the original calculations collected in Topbase.
Datafiles and Grotrian diagrams may be retrieved from the TLUSTY Web
site.\footnote{{\tt http://nova.astro.umd.edu}}

\subsection{Hydrogen and Helium}
\label{hydrogen}

In TLUSTY, v.~201, we have introduced a new default for collisional excitation
in hydrogen. Collisional rates are now computed following \cite{giovanardi87}, 
with expressions valid in a wide temperature domain ($3000\le T\le 500\,000$\,K).
This modification introduces small changes in hydrogen level populations with
repect to the older collisional rates \citep{MHA75}. We compare overlapping
models from the {\sc Ostar2002} and \grid\  grids in \S\ref{GridSect} and we attribute
small changes in Balmer line profiles to differences in hydrogen collisional
excitation. With this exception, hydrogen and helium are treated as described in Paper~I.

\subsection{Carbon}
\label{carbon}

We have added a detailed 40-level model atom for neutral carbon and expanded
the \ion{C}{3} model atom. The new \ion{C}{1} model atom includes all 239 levels
found in Topbase with level energies below the ionization limit. Levels up to
the $4p$ levels ($E~<~83\,000$\,cm$^{-1}$), that is, 13 singlet levels, 14 triplet
levels, and $2s2p^3~^5$S$^0$, are included as individual levels. The 211 levels with
higher excitation are grouped into 12 superlevels, 6 in the singlet system and 6
in the triplet system. All the other quintet levels have energies above the ionization
limit and are unaccounted for in this model atom. 
The original \ion{C}{3} model atom included levels only up to $n=6$. We have examined
in Paper~I the consequences of neglecting more excited levels. While this choice has little
effect on the atmospheric structure, we found that the total C$^{+3}$ to C$^{+2}$
recombination rate is underestimated by a factor 2 to 3. We have therefore opted to
update the \ion{C}{3} model atom and include all levels listed in Topbase (below
the ionization limit). The updated 46-level model atom includes 34 levels with the lower
excitation individually ($E~<~342\,000$\,cm$^{-1}$). Levels with higher excitation
are grouped into 12 superlevels, 6 in the singlet system and 6 in the triplet system.

\subsection{Nitrogen}
\label{nitrogen}

We have added a detailed 34-level model atom for neutral nitrogen and expanded
the \ion{N}{2} and \ion{N}{4} model atoms. The new \ion{N}{1} model atom includes the
lower 27 levels up to $4p~^2$P$^0$ ($E~<~108\,000$\,cm$^{-1}$). The next 88 levels,
up to $n\approx 7$, are grouped into 7 superlevels (3 in the doublet system, 4 in the quartets).
Levels from the sextet system are all above the ionization limit and are neglected.
The original \ion{N}{2} model atom did not incorporate any high levels, $n\ge 6$.
The updated \ion{N}{2} model atom includes 32 individual levels, up to $4s$
($E~<~200\,000$\,cm$^{-1}$), and 3 additional singlets and the four listed
quintet levels below the ionization limit. The remaining 326 levels below the
the limit are grouped into 10 superlevels. The NIST database also lists several
excited levels in intermediate coupling; these levels are skipped because
the OP calculations assumed $LS$-coupling. The \ion{N}{4} model atom was
updated similarly to \ion{C}{3}, including the lower 34 individual levels
($E~<~525\,000$\,cm$^{-1}$), and grouping the 92 levels with higher excitation
($n\le 10$) into 14 superlevels.

\subsection{Oxygen}
\label{oxygen}

The model atmospheres incorporate five ions of oxygen, \ion{O}{1} to \ion{O}{5},
and the ground state of \ion{O}{6}. We have added a detailed 33-level model atom
for neutral oxygen. The new \ion{O}{1} model atom includes the
lower 23 levels up to $5p~^3$P ($E~<~104\,000$\,cm$^{-1}$). All remaining 46 levels,
which are listed in Topbase as being below the ionization limit,
are grouped into 10 superlevels evenly split between the triplet and quintet systems.
\ion{O}{2} and \ion{O}{3} have a relatively rich structure of levels. We have thus
updated these two model atoms to include the high excitation levels. The updated
\ion{O}{2} model atom includes the lower 34 individual doublet and quartet levels,
up to $4p~^2$P$^0$ ($E~<~250\,000$\,cm$^{-1}$), and 2 sextet levels ($3s~^6$S$^0$ and
$3p~^6$P). The higher 182 levels are grouped into 12 superlevels, 6 in the doublet
and 6 in the quartet systems. Levels in intermediate coupling listed in the NIST
database are skipped. The updated \ion{O}{3} model atom includes the first 28
individual levels, up to $3s~^3$P ($E~\le 350\,000$\,cm$^{-1}$), and groups the higher 239
levels below the limit into 13 superlevels in the singlet (5), triplet (5), and quintet (3)
systems. The model atom of \ion{O}{4} is the same as in Paper~I.
Because of lower temperatures, we adopted simplified model atoms for the highest ions. 
\ion{O}{5} model atom only includes the first 6 levels (up to $2p^2~^1$S), and excludes
the higher, very excited levels ($E~>~540\,000$\,cm$^{-1}$). The highest level is the
ground state of \ion{O}{6}, while \ion{O}{7} is neglected altogether. 

\subsection{Neon}
\label{neon}

Simple model atoms have been used in Paper~I. For the \grid\  grid, we have added
a new detailed 35-level model atom for neutral neon, and we have extended the \ion{Ne}{2}
and \ion{Ne}{3} model atoms. The \ion{Ne}{1} model atom includes the lower 23 levels
listed in Topbase, and groups all 168 higher levels into 12 superlevels. Following OP,
this model atom is built assuming $LS$-coupling. However, the level structure of neutral
neon is poorly described in $LS$-coupling, and intermediate $jK$-coupling should be preferred
(e.~g. Sigut 1999). We have constructed a 79-level model atom with fine structure and allowing
for intermediate coupling (Cunha et~al. 2006). Changes in atmospheric structure are negligible
as expected. The predicted \ion{Ne}{1} line strengths in the red spectrum are also little
changed. While an exhaustive differential study of the two model atoms needs to be performed,
we believe thus that the \grid\  model spectra provide a good starting point
to analyze \ion{Ne}{1} lines. Detailed line formation calculations and accurate abundance
determinations would however benefit from using the 79-level model atom. The updated
\ion{Ne}{2} model atom includes the lower 23 individual levels,
up to $4s~^2$P ($E~<~284\,000$\,cm$^{-1}$). The higher 144 levels are grouped into
9 superlevels. The \ion{Ne}{3} model atom includes the first 7 singlet and 10 triplet
levels, up to $3p~^3$P ($E~<~400\,000$\,cm$^{-1}$),
and the first 5 quintet levels individually. All 127 higher levels below the ionization
limit are grouped into 12 superlevels. The \ion{Ne}{4} model atom is identical to the model
used in Paper~I. 

\subsection{Magnesium}
\label{magnesium}

We have added magnesium to the list of explicit species because of the importance
of \ion{Mg}{2} $\lambda\lambda$2798, 2803, 4481 lines as diagnostics in spectral analyses
of B-type stars. The \ion{Mg}{2} model atom includes all
individual levels up to $n=6$ and four superlevels ($7\leq n\leq 10$).
The data extracted from Topbase have been extended to include all levels up to
$10l\,^2$L$^0$. We have included the two fine-structure levels of $3p\,^2$P$^0$ for
an exact treatment of the \ion{Mg}{2} $\lambda$2800 resonance doublet which are
described with depth-dependent, Voigt profiles. Because of the high ionization energy
of Mg$^{2+}$ (about 80\,eV), we have restricted our model of high Mg ions to the
ground state of \ion{Mg}{3} and neglected higher ions.

\subsection{Aluminum}
\label{aluminum}

The UV spectrum of B stars also have important diagnostics lines of aluminum, such as
the resonance lines \ion{Al}{2} $\lambda$1670 and \ion{Al}{3} $\lambda\lambda$1855, 1863,
thus leading us to incorporate aluminum in the \grid\  models. Two ions,
\ion{Al}{2} and \ion{Al}{3}, and the ground state of \ion{Al}{4} are explicitly
considered. The \ion{Al}{2} model atom includes the first 20 levels individually,
up to $5p~^1$P$^0$ ($E~<~126\,000$\,cm$^{-1}$), and groups all 61 higher levels in Topbase
into 9 superlevels. The \ion{Al}{3} was built similarly to the \ion{Mg}{2} model atom.
It includes includes all individual levels up to $n=6$ and four superlevels ($7\leq n\leq 10$).
The data in Topbase have been extended to include all levels up to
$10l\,^2$L$^0$. We have included the two fine-structure levels of $3p\,^2$P$^0$ for
an exact treatment of the \ion{Al}{3} $\lambda$1860 resonance doublet. The UV resonance lines,
\ion{Al}{2} $\lambda$1670 and \ion{Al}{3} $\lambda\lambda$1855, 1863, are described
with depth-dependent, Voigt profiles, while we assume depth-independent Doppler
profiles for all other lines.

\subsection{Silicon}
\label{silicon}

We have added a new \ion{Si}{2} model atom, while keeping the \ion{Si}{3} and
\ion{Si}{4} model atoms used in the {\sc Ostar2002} grid. The \ion{Si}{2} model atom
includes all the levels in Topbase below the ionization limit and 4 levels slightly
above the limit, up to $3p^3~^2$D$^0$ ($E~<~126\,000$\,cm$^{-1}$). Excited doublet
levels are grouped into 4 superlevels ($7\leq n\leq 10$). All other levels are
treated as individual levels. We have included the two fine-structure levels of 
the ground state, $3p\,^2$P$^0$ for an exact treatment of the 6 far-UV resonance
doublets which are described with depth-dependent, Voigt profiles. Autoionization
represents an important channel for the silicon ionization equilibrium. It results
in broad resonances in the OP photoionization cross-sections, and autoionization is thus
incorporated in our models via the OP cross-sections \citep{lanz96}.

\subsection{Sulfur}
\label{sulfur}

The model atmospheres incorporate 4 ions of sulfur, \ion{S}{2} to \ion{S}{5}.
All the model atoms have been updated and extended. We restricted the highest ion
to the ground state of \ion{S}{6} and neglected \ion{S}{7}. The \ion{S}{2} model atom
includes all the energy levels found in Topbase below the ionization limit, except the
3 highly excited levels in the sextet system. The first 23 levels, up to $4s~^2$S
($E~<~137\,000$\,cm$^{-1}$), are included individually. The more excited levels
are grouped into 10 superlevels. All 235 \ion{S}{3} levels in Topbase are incorporated
in the model atom. The lowest 24 singlet and triplet levels, up to $4p~^1$S
($E~<~190\,000$\,cm$^{-1}$), and the first 5 quintet levels are included individually.
The higher levels are grouped into 5 singlet superlevels, 5 triplet superlevels, and
2 quintet superlevels. The \ion{S}{4} model atom accounts for 100 individual levels.
The lowest 21 doublet levels, up to $4p~^2$P ($E~<~293\,000$\,cm$^{-1}$), are
included individually, and more excited doublets are grouped into 3 superlevels.
The model atom also accounts for the fine structure of the \ion{S}{4} ground state,
$3p~^2$P$^0$. The first 11 quartet levels ($E~<~320\,000$\,cm$^{-1}$) are treated
as individual explicit levels, while higher quartet levels are grouped into 2 quartet
superlevels. Finally, the \ion{S}{5} model atom is described with 20 individual
levels, up to $E~<~400\,000$\,cm$^{-1}$; all the higher levels are grouped into 5
superlevels. All lines of the 4 ions are assigned depth-independent Doppler
profiles.

\subsection{Iron}
\label{iron}

We have shifted the selected NLTE iron ions towards lower ionization,
explicitly incorporating 4 ions (\ion{Fe}{2} to \ion{Fe}{5})
in the \grid\  model atmospheres.
The \ion{Fe}{3}, \ion{Fe}{4}, and \ion{Fe}{5} model atoms are the same
models that we used in Paper~I. The \ion{Fe}{2} model atom was built in the
same way and is based on \citet{CD22} extensive semi-empirical calculations. 
Levels of same parity and close energies are grouped into superlevels. The
\ion{Fe}{2} model atom includes 23 even and 13 odd superlevels. The \ion{Fe}{2}
photoionization cross-sections are also extracted from calculations by the
Ohio State group (see Table~\ref{IonTbl}).
They assumed $LS$-coupling, and we could typically assign theoretical
cross-sections to observed levels for the lowest 20 to 30 levels. We have assumed
an hydrogenic approximation for higher-excitation levels. The data are then
combined to setup cross-sections for the superlevels, and resampled using
the Resonance-Averaged Photoionization technique (Bautista et~al. 1998; Paper~I). 
For each ion, we consider the full linelist from \citet{CD22}. TLUSTY dynamically
rejects the weaker lines, based on the line $gf$-values, the excitation energies,
and the ionization fractions. With this selection, mostly \ion{Fe}{2} and \ion{Fe}{3}
lines are included in cooler models, while many more \ion{Fe}{4} and \ion{Fe}{5} lines
are explicitly accounted for in the hotter models. This process selects all the
necessary lines that do contribute to the total opacity,
out of the list of 5.7 million iron lines. Typically, about 1 million lines
are selected. It may select however as few as 500,000 lines and up to 2 million lines
in some models, with the larger numbers in cooler and high gravity models. The adopted
selection criterion is inclusive enough to ensure an appropriate description
of the line blanketing effect which is mainly caused by the strongest 10$^4$--10$^5$
lines \citep{OS02}.


\section{DESCRIPTION OF THE GRID}
\label{GridSect}

The \grid\  grid covers the parameter space of early B-type stars in a dense
and comprehensive way. We have selected 16 effective temperatures,
$15\,000$\,K$\leq$\teff $\leq 30\,000$\,K, with 1\,000\,K steps, 13 surface
gravities, $1.75\leq\log g\leq 4.75$, with 0.25\,dex steps, and 6 chemical
compositions, from twice to one-tenth the solar metallicity and metal-free models.
The effective temperature range covers spectral types B0 to B5 \citep{SK82}.
The lower limit in $\log g$ is set by the Eddington limit (see \S\ref{GradSect}). 
Solar abundances refer to \citet[][Sun98]{Sun98}. We have assumed a solar helium
abundance, He/H=0.1 by number. All other chemical abundances are scaled from
the solar values. The microturbulent velocity was set to \vtur = 2\,\kms.
On the main sequence, the covered temperature range corresponds to stars with
initial masses between 4 and 15 solar masses. At lower gravities, the \grid\  grid
covers the case of more massive stars evolving off the main sequence.

Recent spectroscopic studies reveal that early B supergiants show strongly
processed material at their surface \citep[e.~g.,][]{dufton05, crowther06}.
In these NLTE analyses, large microturbulent velocities (\vtur~$\geq 10$\,\kms)
are derived for most stars. We emphasize here that these high microturbulences
do not result from neglecting the wind and from assuming hydrostatic equilibrium.
Like the case of O stars, such high microturbulent velocities are obtained both
from TLUSTY models and from unified models. Therefore, we decided to supplement
the initial grid with two additional sets of model atmospheres suitable for
supergiants ($\log g\leq 3.0$). In the first set, we adopted \vtur = 10\,\kms,
but kept solar or solar-scaled abundances. In the second set, in addition
to the larger microturbulence, we increased the helium abundance to He/H=0.2
by number, increased the nitrogen abundance by a factor of 5, and halved the carbon
abundance (hence, an order of magnitude increase in the N/C abundance ratio,
reflecting CNO-cycle processed material brought to the stellar surface).

\begin{figure}

\includegraphics[scale=0.7,angle=90]{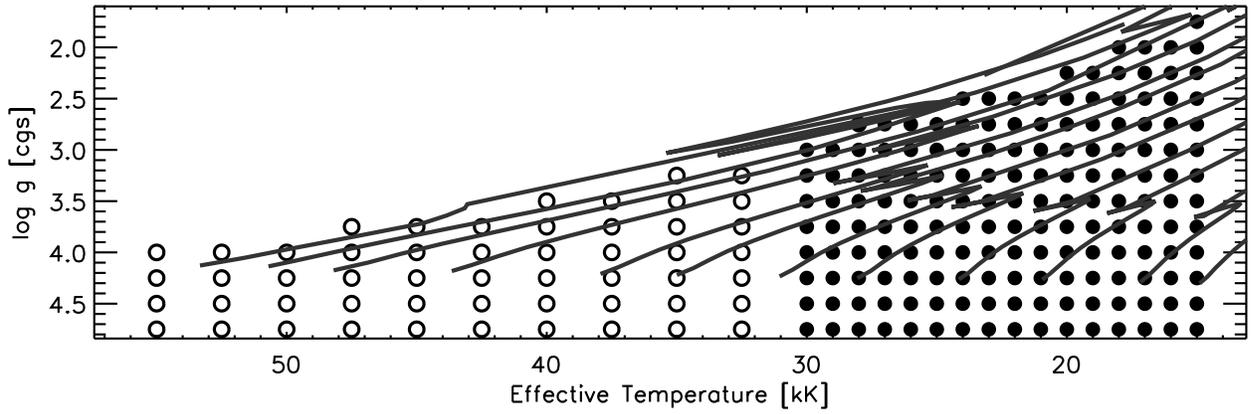}
\figurenum{1}
\caption{Selected \grid\  grid points (full) and {\sc Ostar2002} grid points (open)
in the $\log g\  vs.$ \teff\  plan. The two grids overlap at \teff = 30\,000\,K.
Geneva evolutionary tracks \citep{Geneva1} are shown for solar metallicity, and correspond
to models with initial masses
of 120, 85, 60, 40, 25, 20, 15, 12, 9, 7, 5, and 4\,$M_\sun$ from left to right, respectively.
\label{TeflogFig}}
\end{figure}

For each of the five metallicities, we have thus computed 265 model atmospheres.
This includes a full set of 163 models at the lower microturbulence, and two sets
of 51 models at the higher microturbulence (include only low gravity models).
At the highest metallicities,  a few models
very close to the Eddington limit could not be converged because of physical and
numerical instabilities; these few models are thus skipped.
Fig.~\ref{TeflogFig} illustrates the sampling of the $\log g$~vs. \teff\  diagram
by our two grids of NLTE model atmospheres. The two grids overlap at \teff~= 30\,000\,K,
and we show below a comparison. 

\subsection{Output Products and Availability}

The model atmospheres are available at the TLUSTY Web
site.\footnote{\tt http://nova.astro.umd.edu}
Similarly to the {\sc Ostar2002} grid,
each model is characterized by a unique filename describing the parameters
of the model, for example BS25000g275v10CN. The first letters indicate the
composition, followed by the effective temperature,
the gravity and the turbulent velocity. For models with altered surface
composition (supplemental set~2), ``CN'' is appended to the filename.
We have adopted the same key for the models' overall composition as in the
{\sc Ostar2002} grid
(see Table~\ref{KeyTbl}), that is: twice solar (``BC
models''), solar (``BG models''), half solar (``BL models''),  one fifth solar
(``BS models''), one tenth solar (``BT models''),
and metal-free (``BZ models'').\footnote{Model ``BS25000g275v10CN'' thus
corresponds to a model with \teff~=25\,000\,K, $\log g = 2.75$,
\vtur~=10\,\kms, one fifth solar metallicity, and altered He, C, and N surface abundances.}
Model atmospheres with even lower metallicities as in the {\sc Ostar2002} grid may
be made available at a future time.

Each model comes as a set of six files,
with an identical filename's root but a different extension. A complete description
of the files' content and format can be found in TLUSTY  User's guide (see
TLUSTY web site); for reference, we describe them only very briefly here: \\ [1.5mm]
{\tt model.5}: General input data; \\
{\tt model.nst}: Optional keywords; \\[1.5mm]
{\tt model.7}: Model atmosphere: Temperature, electron density, total density
                and NLTE populations as function of depth; \\[1.5mm]
{\tt model.11}: Model atmosphere summary; \\
{\tt model.12}: Model atmosphere: Similar to {\tt model.7}, but NLTE populations
                 are replaced by NLTE $b$-factors (LTE departure coefficients); \\[1.5mm]
{\tt model.flux}: Model flux distribution from the soft X-ray to the far-infrared
                    given as the Eddington flux\footnote{The flux at
                    the stellar surface is $F_\nu = 4\pi\,H_\nu$.}
                    $H_\nu$ [in erg\,s$^{-1}$\,cm$^{-2}$\,Hz$^{-1}$] as function of
                    frequency.
                    
The model fluxes are provided at all frequency points included in the calculations
(about 380\,000 points with an irregular sampling for the low microturbulence models;
about 175\,000 points for the ``v10'' models). Additionally,
we have calculated detailed emergent spectra with SYNSPEC, version~48, in the
ultraviolet ($\lambda\lambda$900-3200\,\AA) and in the optical
($\lambda\lambda$3200-10\,000\,\AA). Filename extensions are {\tt model.uv.7},
{\tt model.vis.7}, respectively, and {\tt *.17} for the continuum spectra.
Additional spectra in other wavelength ranges,
with altered chemical compositions, or with different values of the microturbulent
velocity can be readily computed using SYNSPEC. The spectrum synthesis
requires three input files, {\tt model.5, model.7,} and {\tt model.nst},
and the necessary atomic data files (model atoms and the relevant linelist).

The analysis of individual stars may require to interpolate within the model
grid. Interpolation procedures were discussed in Paper~I, \S5.2. Since
the \grid\  grid is sampled with a temperature step (1000\,K) that is even finer than
the step in the {\sc Ostar2002} grid, these procedures can be applied. As the
safest method, however, we advise to interpolate the atmospheric structure for the
selected parameters, followed by recalculating the spectrum with SYNSPEC.
This approach provides a very fast way to apply these fully-blanketed NLTE model
atmospheres to detailed spectrum analyses of B stars.

\subsection{Model Sensitivity}

We have investigated the effect of several assumptions made to calculate the \grid\ 
model atmospheres, including the treatment of iron opacity, the choice of the
``standard'' solar abundances, and the adopted model atoms. For these tests,
we have selected the solar composition model, \teff = 25\,000\,K,
$\log g~=~4.0$, \vtur~=~2\,\kms, and one solar composition model (\teff = 30\,000\,K,
$\log g~=~3.0$) where the two grids overlap.

\begin{figure}
\includegraphics[scale=0.79,angle=0]{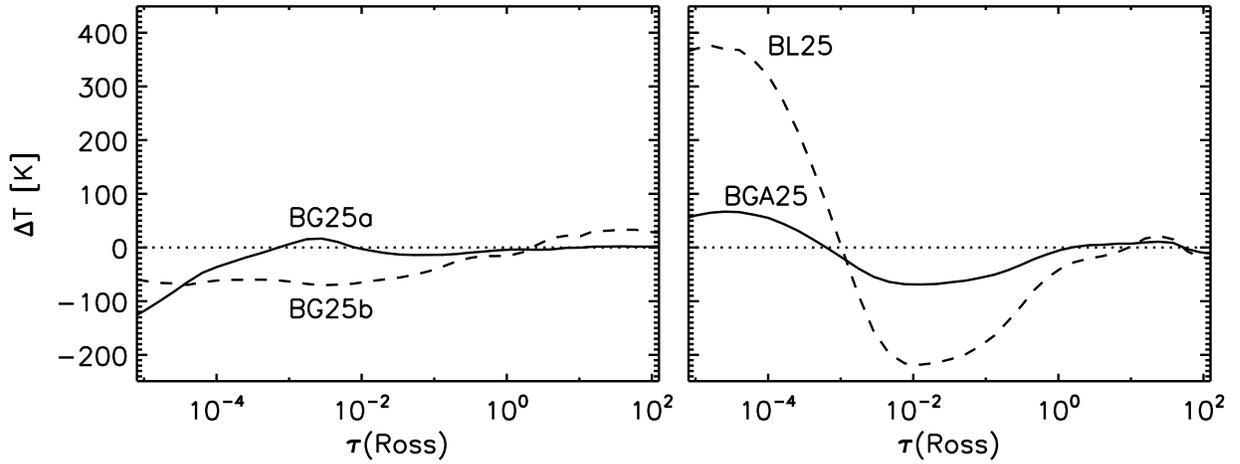}
\figurenum{2}
\caption{Change in the atmospheric temperature structure with respect to the
reference model BG25000g400v2 (\teff~=~25\,000\,K, $\log g=4.0$, solar composition,
and \vtur~=~2\,\kms). The left panel shows the changes due to different assumptions
in Opacity Sampling, using larger frequency sampling steps (BG25a, full line) or
only selecting the strongest \ion{Fe}{2} to \ion{Fe}{5} lines (BG25b, dashed line).
The right panel shows the effect of the assumed abundances, with lower CNO abundances
and higher Ne abundance (BGA25, full line), or lowering the overall metallicity by half
(BL25, dashed line).
\label{OSFig}}
\end{figure}

The first test expands at lower \teff\  on the tests of iron line-blanketing made
in Paper~I, \S8. Based on earlier experience and mostly on the importance of
representing blends as accurately as possible,\footnote{see \cite{najarro06} for an
example of the importance of \ion{Fe}{4} lines blending \ion{He}{1}~$\lambda$584.}
we preferentially adopt Opacity Sampling (OS) over the Opacity Distribution
Function approach. In order to ensure the best description of line blends, we have
used a small frequency sampling step (0.75 fiducial Doppler width) as the standard.
We have then tested the resulting effect on the atmospheric structure
of a larger sampling step and of a different line strength selection criterion.
Compared to the reference solar composition model BG25000g400v2,
model BG25a uses a frequency sampling step that is 40 times larger
for the iron lines, without any changes for lines of lighter species, hence reducing
the total number of frequencies by a factor of about 3. Model BG25b limits the dynamically
selected iron lines to the strongest lines, selecting about 85,000 lines of \ion{Fe}{2}
to \ion{Fe}{5} compared to over 1,500,000 lines in the reference model. Fig.~\ref{OSFig}
(left panel) displays the change in the atmospheric temperature structure with repect
to the reference model. The temperature differences remain small, quite comparable
to the changes seen in similar tests for O star model atmospheres. As a reference,
the typical numerical accuracy achieved in conserving the total radiative flux corresponds
to uncertainties up to 10~--~20\,K on the local temperature. The limited extent of temperature
changes found in these tests implies only minimal changes in the predicted spectra,
as exemplified by the test with different standard abundances where the temperature
changes are somewhat larger (see below, and Figs.~\ref{OSFig} and~\ref{ABFig}). 
We believe that the tests that we performed for O and B star model atmospheres are
representative to the numerical accuracy of Opacity Sampling in hot model atmospheres.
The larger sampling step might thus be quite sufficient for most purposes, hence allowing
a substantial saving in computer time (up to a factor of 3). However, because we intend
the \grid\  grid to serve as reference NLTE model atmospheres, and since accuracy testing
is practicable only in a limited extent, we have elected to keep the finest sampling to
ensure the best description of iron line blanketing.

When we initiated the work on the {\sc Ostar2002} and \grid\  grids, we adopted the
standard solar abundances from \cite{Sun98}. In the interim, several studies showed
that the carbon, nitrogen, and oxygen abundances in the Sun are significantly lower,
about 70\% of the earlier standard values \citep{AGS05}. The
new lower abundances are in good agreement with abundances measured in B stars in 
the solar neighborhood \citep[e.~g.,][]{cunha94}. Additionally, \cite{cunha06} found
that the neon abundance in the Orion B stars is twice the standard value, and they argued
that this value might be used as a good proxy for the solar abundance. Accordingly, we
investigated the differences in atmospheric structure and predicted spectrum from these
updated C, N, O, and Ne abundances, calculating a model atmosphere using \teff~=~25\,000\,K,
$\log g=4.0$, \vtur~=~2\,\kms, and the newer abundances. For the comparison, we considered
4 different model atmospheres: two models in the grid for solar (BG25) and half-solar
(BL25) metallicities
(since the new CNO values fall in between), the BGA25 model atmosphere with the new abundances,
and a model spectrum BGZ25 calculated with the solar composition atmospheric structure
and applying the new abundances in the spectrum calculation step only. Fig.~\ref{OSFig}
(right panel) shows the temperature differences between the first three models. The changes
resulting from the changes in the light element abundances remain limited 
($\Delta T~<~90$\,K), and they are as much as 4 times smaller than the differences between
the solar and half-solar metallicity models. This comparison shows the importance of iron
line blanketing in establishing the temperature structure. We then compare the resulting
effect of the abundance changes on the predicted spectra. We show in Fig.~\ref{ABFig} the
spectral region around H$\gamma$, also including \ion{He}{1}~$\lambda$4388 and many
\ion{O}{2} lines, and the relative differences between the model spectra. The absolute
{\em continuum} flux level changes by less than 1\%, and up to 2\% for the half-solar metallicity
model. The largest changes, 4\% and higher, are seen in {\em line cores}, mostly in
the strength of \ion{O}{2} lines which are directly related to
changes in the oxygen abundance in the spectrum
calculation. Most importantly, however, the BGZ25 model spectrum 
is very similar to the BGA25 model spectrum (see Fig.~\ref{ABFig}, middle panel of 3
showing the relative changes) and, therefore, shows that the largest 
spectrum changes directly result indeed from abundance changes in the spectrum synthesis step
rather than from the indirect change in the atmospheric structure. In most cases,
it might therefore be appropriate to use the solar composition
model atmospheres and only to recalculate the spectra with the updated abundances of light
species. 
We finally stress that the highest accuracy
will always be reached when tailoring the model atmospheres to the stars studied, using the
closest estimates of all stellar parameters and abundances; the \grid\  grid 
primarily intends to provide the best starting point for these detailed analyses.

Finally, we examine the sensitivity of model stellar atmospheres to the adopted model
atoms by comparing a model atmosphere with identical parameters (\teff~=~30\,000\,K,
$\log g=3.0$, solar composition, and \vtur~=~10\,\kms) from the two grids, \grid\ 
and {\sc Ostar2002}. Sect.~\ref{AtDataSect} details the atom models and changes between
the two grids. In summary, the newer models includes lower ionization ions as well as
more complete model atoms, in particular for \ion{C}{3}, \ion{N}{2}, \ion{N}{4}, \ion{O}{2},
\ion{O}{3}, and the Ne and S ions. Fig.~\ref{O2B6Fig} illustrates the very high consistency
between the two grids: the very rich ultraviolet and the visible spectra predicted by
these two calculations are very much the same. Rotational broadening ($V \sin i~=~50$\,\kms)
has been applied to the spectra to show a comparison that is more relevant to actual
spectral analyses. In the visible spectrum, we particularly note on one hand the excellent
agreement in H$\gamma$, \ion{He}{1}~$\lambda$4471, and \ion{He}{1}~$\lambda$4388 in emission,
while weak metal lines on the other hand reveal some discrepancies. A closer
examination shows that Balmer lines are predicted very slightly weaker in the \grid\  models
than in the {\sc Ostar2002} models, which we interpret as resulting from
the adoption of different bound-bound collisional rates in hydrogen. This small difference
has an insignificant impact on the derived stellar parameters. All metal lines showing
changes in the optical spectrum arise from highly-excited levels.
Hence, these differences are likely a
consequence of the newer model atoms that explicitly include higher levels in the
NLTE calculations. The $gf$-values of some lines in the linelist used by SYNSPEC have also
been updated from the original Kurucz data to OP data for consistency between the model
atmosphere and the spectrum synthesis calculations. We expect the OP values to be of
higher accuracy in general. From this last test, we may therefore conclude that
the two model atmosphere grids will yield consistent results. Although these models
include a detailed and extensive treatment of line blanketing, we stress that these
models should not be applied blindly to the analysis of some weak lines in the visible
and infrared range without a line-by-line assessment of the adequacy of the model
atoms in each case.


\section{REPRESENTATIVE RESULTS}
\label{ResSect}

In this section, we show some basic properties and important trends describing
model atmospheres of B-type stars, in particular a comparison to Kurucz (1993)
LTE model atmospheres and spectra, ionization structure, bolometric corrections,
and radiative accelerations.

\subsection{Comparison to LTE Model Atmospheres}
\label{KuruSect}

We first compare the atmospheric structure of NLTE \grid\  and LTE Kurucz (1993)
model atmospheres. Figure~\ref{TtauFig} displays the temperature stratification of a series
of solar composition and one-tenth solar metallicity NLTE and LTE models, \teff~=~25\,000\,K,
$\log g~=$ 3.0 and 4.0. At large depths ($\tau_{\rm Ross}\ga 1$, corresponding to mass column
densities larger than 0.5\,g\,cm$^{-2}$) where the departures from LTE are small,
the LTE and NLTE atmospheric temperature structures are very similar. This agreement
thus provides good support to the two independent approaches (Kurucz ODFs and our OS treatment)
to incorporate the opacity of millions of lines in model stellar atmospheres. 
In shallower layers, the local temperature in NLTE models is higher than in the
LTE models. In the NLTE models, temperature is basically determined by a balance
between heating by the Balmer hydrogen lines and cooling by lines of heavy elements
(carbon and heavier elements). Figure~\ref{TtauFig} illustrates that the classical
NLTE temperature rise is not completely removed by the cooling from metal lines, even
at solar metallicity, and shows the metallicity dependence of this effect. We may
therefore expect from this comparison that LTE and NLTE {\em continuum} spectra
will not differ too much, while the core of strong lines and lines from minor ions will
be most affected by departures from LTE.

We then compare the predicted spectra of our NLTE line-blanketed model atmospheres with
\cite{CD13} LTE model spectra. Fig.~\ref{FKurFig} displays a comparison of the
predicted LTE and NLTE spectral energy distributions for 3 solar composition models
with $\log g = 3.0$ and \vtur~=~2\,\kms. For this comparison, we use the spectra
directly calculated by TLUSTY (files {\tt model.flux} described earlier), but
not detailed spectra computed by SYNSPEC. For clarity, the spectra
are smoothed over 800 frequency points; this roughly simulates the 
10\,\AA~resolution of \citet{CD13} model spectra. This figure reveals some
differences in the continuum fluxes, most noticeably in the near ultraviolet
where the LTE fluxes are about 10\% higher than the NLTE predictions. The lower NLTE
fluxes result from the overpopulation of the \ion{H}{1}~$n=2$ level at the depth
of formation of the continuum flux, hence implying
a larger Balmer continuum opacity. A smaller difference is seen in the Paschen
continuum of the cooler models because of overpopulation of the $n=3$ level.
At higher surface gravities, these differences are still present albeit reduced.
To conserve the total flux, the LTE models show a lower flux in the far and extreme
ultraviolet. In the hottest model shown (\teff~=~25\,000\,K), the LTE prediction in
the Lyman continuum in a factor of 2 lower than the NLTE flux. In the cooler
models, the LTE flux is lower at wavelengths shorter than Ly~$\alpha$.

In the last step comparing the \grid\  models to LTE models,
we examine detailed line profiles calculated with SYNSPEC, using either a TLUSTY
model atmosphere or a Kurucz model atmosphere (\teff~=~20\,000\,K, $\log g=3.0$,
solar composition, and \vtur~=~2\,\kms). We have convolved the spectra with
a rotation broadening (10\,\kms) and normalized the spectra to the continuum.
To compute the spectra, we have used the same atomic line list, hence all
differences result from level populations departing from the LTE value
or differences in the atmospheric model structure.
Some typical behaviors are illustrated in Fig.~\ref{LinesFig}. In the NLTE model,
the hydrogen Balmer lines are broader and stronger. For the main part, this is
the result of the overpopulation of the $n=2$ level. Stellar analyses relying
on LTE model atmospheres therefore tend to overestimate surface gravities
derived from Balmer line wings. Other optical lines differ little (e.~g., \ion{O}{2}
lines, or \ion{Al}{3}~$\lambda$4480) or are somewhat stronger in the NLTE
model spectrum (e.~g., \ion{He}{1}~$\lambda$4471, \ion{Mg}{2}~$\lambda$4481, or
\ion{Si}{3}~$\lambda\lambda$4552, 4567, 4575). In the latter case, this implies that
the abundance of some species might be overestimated from LTE predictions; for instance,
a LTE analysis of \ion{Ne}{1} lines overestimates the neon abundance up to 0.5~dex
\citep{cunha06}. Finally, Fig.~\ref{LinesFig} shows a comparison in the far ultraviolet
around 1300\,\AA, illustrating the effect of NLTE radiative overionization. Lines
of the dominant ions, such as \ion{Si}{3}~$\lambda\lambda$1301, 1303, remain virtually
unchanged, while lines of minor ions (e.~g., \ion{Si}{2}~$\lambda\lambda$1305, 1309)
are significantly weaker in the NLTE model. The weak \ion{Si}{2} lines mainly result
from the typical radiative overionization found in NLTE models; differences in
atmospheric structure have a limited influence. While using Kurucz LTE models might
have been a reasonable choice for analyzing UV and optical spectra of B-type stars,
these comparisons demonstrate that the new \grid\  NLTE line-blanketed model
atmospheres represent a significant progress for determining the stellar parameters
and the surface chemical composition of these stars. We plan to present a systematic
study of various effects and differences between LTE and NLTE models,
and their influence on deduced stellar parameters, in a future paper.

\subsection{Ionization Fractions}
\label{IonSect}

Figures~\ref{IonHeFig}--\ref{IonFeFig} display the ionization fractions of all
explicit species (with the exception of hydrogen) in model atmospheres spanning
the range of temperature of the \grid\  grid. For completion, we have added two
hotter O star model atmospheres (\teff\ = 40,000\,K and 50,000\,K), thus providing
a comprehensive view of the change in ionization from 15\,000\,K
to 50\,000\,K.\footnote{Fig.~7 of Paper~I displaying the silicon ionization
fraction in O stars was unfortunately drawn incorrectly; Fig.~\ref{IonSiFig} shows
the correct silicon ionization.} We show
the ionization structure of solar composition, main-sequence ($\log g=4.0$) 
star atmospheres. The 6 selected model atmospheres roughly correspond to spectral
types B5, B2.5, B1, B0, O5, and O2-O3. These figures thus give a straightforward
illustration of the change in the expected line strength for various ions along
the spectral type sequence, and support our choice of explicit NLTE ions. At depth
($\tau_{\rm Ross}\ga 10$), the ionization fractions may become incorrect when higher
ions have been neglected, most particularly for magnesium and aluminum. We
stress however that this restriction does not affect the predicted spectra at all.
Fig.~\ref{IonGravFig} compares the oxygen and iron ionization in model atmospheres
with increasing surface gravities. At \teff~=~20\,000\,K, the ionization shifts
up by one degree between a main-sequence model ($\log g = 4.25$) and a supergiant
model ($\log g = 2.25$). Higher ionization at lower gravities is a well-known result,
which follows from the Saha formula and from the typical NLTE radiative overionization.
Finally, a comparison of the ionization structure of models with low and high microturbulent
velocities shows that ionization is very slightly higher in models with the higher
microturbulence. There is virtually no change in the dominant ion; with respect to
this ion, lower ions are less populated and higher ions are more populated in models
with the higher microturbulence. We interpret this change as a consequence of the
stronger line blanketing. The enhanced line blocking results in slightly higher
temperature at depth because of backwarming. The ions are thus exposed to a harder
radiation field, hence the slight ionization shift.

These ionization fractions are useful to roughly assess the expected
strengths of individual spectral lines. However, the ionization fractions should
not be  taken too literally. In some important cases, lines may originate
from very high-lying levels whose populations may be a tiny fraction
of the total abundance of a specific ion. In those cases, relying on
the ionization fractions displayed in Figs~\ref{IonHeFig}--\ref{IonFeFig}
alone may be misleading. The interested reader may however always access 
the individual level populations that are stored in files {\tt model.7},
as well as the individual NLTE departure coefficients stored
in files {\tt model.12} to estimate better the expected line
strengths, or the ratios of line strengths for lines of
different ions of the same species.

\subsection{Bolometric Corrections and Ionizing Fluxes}
\label{BCSect}

Bolometric corrections (Table~\ref{BCTbl}) are calculated using the following expression:
\begin{equation}
BC_V = m_{\rm bol} - V = (-2.5\log F_{\rm bol} - 11.487) - (-2.5\log F_V - 21.100)
\end{equation}
where $F_{\rm bol}$ is the bolometric flux and $F_V$ is the flux through
the Johnson $V$ filter, computed from the model atmospheres' flux distribution.
The bolometric flux is computed by trapezoidal integration over the complete frequency
range, while we have used an IDL version of the program {\sc ubvbuser} \citep{CD13}
kindly made available by Wayne Landsman to obtain the $V$ magnitudes.  The first constant
is defined by assuming a solar  value, $BC_V^\sun = -0.07$, while the second constant
defines the zero point of the $V$ magnitude scale.

Fig.~\ref{BCFig} illustrates the major dependence of the bolometric correction with
the effective temperature (solar metallicity, $\log g=4.0$). We have extended the plot
to encompass both the \grid\  and the {\sc Ostar2002} models. Dependences in terms
of the other stellar parameters (gravity, metallicity and microturbulent velocity) are
much weaker than the temperature dependence. While we find differences up to 0.1~mag
over the whole range in metallicity (twice solar to metal-free), the changes remain
limited to a few hundredths of a magnitude for different gravities and microturbulences.
Larger differences are found for models very close to the Eddington limit,
but these models are in a  domain where our basic assumption of static atmospheres starts
to break up. The small step (0.05~dex) in the overlapping
range between the two grids is therefore most likely a consequence of the differences
in the treatment of opacities, that is, it likely results from the different ions, levels
and lines explicitly included in the model atmospheres. 
The bolometric corrections of ``CN'' models do not differ from models with unaltered
compositions, and they are thus not listed in Table~\ref{BCTbl}.

Ionizing photons are emitted essentially by hot, massive stars, and
the extreme ultraviolet flux drops precipitously in early B stars with diminishing
effective temperatures. Therefore, we only provide in Table~\ref{Q0Tbl}
a summarized extension of the {\sc Ostar2002} Lyman continuum fluxes to the early B star
domain.

\subsection{Radiative Acceleration and Effective gravities}
\label{GradSect}

The comprehensive treatment of opacities in the \grid\  model atmospheres allows
for an accurate estimate of radiative pressure on the atmospheric structure. TLUSTY
thus provides a run of radiative acceleration with depth, see Fig.~2 in Paper~I for
an illustration. In every model atmosphere, the radiative acceleration goes through a
local maximum in the continuum forming region ($\tau_{\rm Ross}\approx 1$), and 
then shows another strong
increase in very superficial layers ($\tau_{\rm Ross}\la 10^{-6}$). In low gravity
models, the radiative acceleration may exceed gravity at these depths indicating
that the top layers are unstable. The assumption of hydrostatic
equilibrium then break down there. We have numerically limited the radiative acceleration
in these superficial layers to ensure convergence following the prescription used
in Paper~I. The model spectra are not affected
by this approximation, because these superficial layers only influence the strong
resonance lines that form in a stellar wind. The photosperic stability is more
appropriately defined by the maximum reached by the radiative acceleration around
optical depth unity (see Fig.~2, Paper~I). Fig.~\ref{EddingFig} displays isocontours
of this maximum relative to the gravitational acceleration,
$\Gamma_{\rm rad} = \max(g_{\rm rad})/g$, as a function of effective temperature and
gravity. Models tend to become numerically unstable when
$\Gamma_{\rm rad} > 0.9$, setting thus the gravity limit in our grid.

We can extrapolate these radiative acceleration to define the limit of photospheric
stability when $\Gamma_{\rm rad}$ reaches unity, thus providing with an estimate
of the Eddington limit in absence of rotation (see Fig.~\ref{EddingFig}).
Table~\ref{EffGravTbl} gives $\log g_{\rm Edd}$, the Eddington limit in term
of $\log g$, as a function of effective temperature from 55\,000 to 15\,000\,K.
We have carried out the same extrapolation with metal-free model atmospheres, and
found that the corresponding Eddington limit is reached about 0.05~dex lower than
the values derived for solar composition models. \cite{lamers88} performed
a similar exercise, defining the Eddington limit by extrapolating the radiative
accelerations calculated with Kurucz model atmospheres. Their estimates agree
very well with our results over the whole range of temperature.

Finally, we can derive effective gravities, $g_{\rm eff} = g - \max(g_{\rm rad})$,
in absence of rotation (see Table~\ref{EffGravTbl}). These effective gravities may be
used to estimate the characteristic pressure scale height ($h = P_{\rm gas}/g_{\rm eff}$)
in the photosphere, after possibly correcting $g_{\rm eff}$ further for centrifugal
acceleration. These scale heights indicate that the photosphere of main-sequence B
stars are compact with scale heights typically smaller than 1\% of the stellar radius.
On the other hand, the photosphere of extreme B supergiants ($\log g\le 2.0$) are quite extended
($h > 1 R_\sun$, or several percent of their radius).
Caution should be exercised when applying these model atmospheres
to analyze the spectra of these supergiants.


\section{CONCLUSION}
\label{ConclSect}

We have constructed a comprehensive grid of metal line-blanketed, NLTE,
plane-parallel, hydrostatic model atmospheres for the basic parameters
appropriate to early B-type stars.
The \grid\  grid considers 16 values of effective temperatures 
15\,000\,K $\leq$\teff\ $\leq$ 30\,000\,K, with 1\,000\,K steps, 
13 surface gravities, $1.75\leq\log g\leq 4.75$ with  
0.25\,dex steps, and 6 chemical compositions, from metal-rich 
relative to the Sun to metal-free.  The lower limit of
$\log g$ for a given \teff\  is actually set by an approximate 
location of the Eddington limit. 
The complete \grid\  grid is available at our website 
at {\tt http://nova.astro.umd.edu}.

We have intended to provide 
a more or less definitive grid of model atmospheres in the context of
one-dimensional, plane-parallel, homogeneous, hydrostatic models
in radiative equilibrium.
We have attempted to take into account essentially all important
opacity sources (lines and continua) of all astrophysically important
ions. Likewise, we have attempted to consider all relevant atomic 
processes that determine the excitation and ionization balance of
all such atoms and ions. The models explicitly include, and allow
for departures from LTE, 46 ions of H, He, C, N, O, Ne, Mg, Al, Si, S,
and Fe, and about 53\,000 individual
atomic levels (including about 49\,300 iron levels) grouped into
1127 superlevels. Line opacity includes about 39\,000 lines from
the light elements and 500\,000 to 2 million iron lines dynamically
selected from a list of about 5.65 million lines. 

Although we spent all effort to make sure that the treatment
of atomic physics and opacities is as complete and accurate
as possible, there are still several points we are aware of that 
were crudely approximated.  Those approximations were necessitated
not by shortcomings of our modeling scheme or a lack of adequate
computer resources, but by the present lack of sufficient atomic data.
These approximations include a crude and approximate treatment of
collision rates, a neglect of very high energy levels of light metals,
and ignoring charge exchange reactions. Despite our updating most model
atoms to include levels up to $n=10$, we caution that the analysis of
weak lines from highly excited energy levels in the optical and infrared
spectrum might require to construct even more detailed model atoms,
supplementing the model atoms with higher excitation levels, or
possibly also splitting some superlevels into individual levels.

Despite the remaining approximations and uncertainties,
we believe that the \grid\  grid represents a definite improvement
over previous grids of B star model atmospheres. Combined with
our earlier {\sc Ostar2002} grid of O star model atmospheres,
we hope that these new model atmospheres will thus be useful for a number
of  years to come for analyzing the spectrum individual O and B stars,
as well as constructing composite model spectra of clusters of young massive
stars, OB associations, and starburst galaxies.

\acknowledgments

This work was supported by NASA grant NAG5-13214
(FUSE D162 program), NASA ADP grants, and by several grants (GO 9116,
GO 9848) from the Space Telescope Science Institute, which is operated
by the Association of Universities for Research in Astronomy, Inc., under NASA
contract NAS5-26555.


\singlespace
\clearpage

\begin{deluxetable}{lrrrc}
\tabletypesize{\small}
\tablecaption{Atomic data included in the NLTE model atmospheres. \label{IonTbl}}
\tablehead{
\colhead{Ion} & \colhead{(Super)Levels}   & \colhead{Indiv. Levels}   &
\colhead{Lines} & \colhead{References}
}
\startdata
\ion{H}{1}  &   9\phm{Levels}  &  80\phm{Levels}  &     172\phm{a}  &                \\*
\ion{H}{2}  &   1\phm{Levels}  &   1\phm{Levels}  &     \nodata     &                \\
&&&&\\
\ion{He}{1} &  24\phm{Levels}  &  72\phm{Levels}  &     784\phm{a}  & 1  \\*
\ion{He}{2} &  20\phm{Levels}  &  20\phm{Levels}  &     190\phm{a}  &                \\*
\ion{He}{3} &   1\phm{Levels}  &   1\phm{Levels}  &     \nodata     &                \\
&&&&\\
\ion{C}{1}  &  40\phm{Levels}  &  239\phm{Levels}  &    3201\phm{a} & 2 \\*
\ion{C}{2}  &  22\phm{Levels}  &  44\phm{Levels}  &     238\phm{a}  & 3 \\*
\ion{C}{3}  &  46\phm{Levels}  &  95\phm{Levels}  &     738\phm{a}  & 4 \\*
\ion{C}{4}  &  25\phm{Levels}  &  55\phm{Levels}  &     330\phm{a}  & 5  \\*
\ion{C}{5}  &   1\phm{Levels}  &   1\phm{Levels}  &     \nodata     &                \\
&&&&\\
\ion{N}{1}  &  34\phm{Levels}  & 115\phm{Levels}  &     785\phm{a}  & 6 \\*
\ion{N}{2}  &  42\phm{Levels}  & 247\phm{Levels}  &    3396\phm{a}  & 2 \\*
\ion{N}{3}  &  32\phm{Levels}  &  68\phm{Levels}  &     549\phm{a}  & 3 \\*
\ion{N}{4}  &  48\phm{Levels}  & 126\phm{Levels}  &    1093\phm{a}  & 4 \\*
\ion{N}{5}  &  16\phm{Levels}  &  55\phm{Levels}  &     330\phm{a}  & 5  \\*
\ion{N}{6}  &   1\phm{Levels}  &   1\phm{Levels}  &     \nodata     &                \\
&&&&\\
\ion{O}{1}  &  33\phm{Levels}  &  69\phm{Levels}  &     418\phm{a}  & 7 \\*
\ion{O}{2}  &  48\phm{Levels}  & 218\phm{Levels}  &    3484\phm{a}  & 6 \\*
\ion{O}{3}  &  41\phm{Levels}  & 267\phm{Levels}  &    3855\phm{a}  & 2 \\*
\ion{O}{4}  &  39\phm{Levels}  &  94\phm{Levels}  &     922\phm{a}  & 3 \\*
\ion{O}{5}  &   6\phm{Levels}  &   6\phm{Levels}  &       4\phm{a}  & 4  \\*
\ion{O}{6}  &   1\phm{Levels}  &   1\phm{Levels}  &     \nodata     &                \\
&&&&\\
\ion{Ne}{1} &  35\phm{Levels}  & 191\phm{Levels}  &    2715\phm{a}  & 8 \\*
\ion{Ne}{2} &  32\phm{Levels}  & 167\phm{Levels}  &    2301\phm{a}  & 7 \\*
\ion{Ne}{3} &  34\phm{Levels}  & 149\phm{Levels}  &    1354\phm{a}  & 7 \\*
\ion{Ne}{4} &  12\phm{Levels}  &  18\phm{Levels}  &      38\phm{a}  & 6 \\*
\ion{Ne}{5} &   1\phm{Levels}  &   1\phm{Levels}  &     \nodata     &                \\
&&&&\\
\ion{Mg}{2} &  25\phm{Levels}  &  53\phm{Levels}  &     306\phm{a}  & 9 \\*
\ion{Mg}{3} &   1\phm{Levels}  &   1\phm{Levels}  &     \nodata     &                \\
&&&&\\
\ion{Al}{2} &  29\phm{Levels}  &  81\phm{Levels}  &     536\phm{a}  & 10 \\*
\ion{Al}{3} &  23\phm{Levels}  &  53\phm{Levels}  &     306\phm{a}  & 9 \\*
\ion{Al}{4} &   1\phm{Levels}  &   1\phm{Levels}  &     \nodata     &                \\
&&&&\\
\ion{Si}{2} &  40\phm{Levels}  &  64\phm{Levels}  &     392\phm{a}  & 11 \\*
\ion{Si}{3} &  30\phm{Levels}  & 105\phm{Levels}  &     747\phm{a}  & 10 \\*
\ion{Si}{4} &  23\phm{Levels}  &  53\phm{Levels}  &     306\phm{a}  & 9 \\*
\ion{Si}{5} &   1\phm{Levels}  &   1\phm{Levels}  &     \nodata     &                \\
&&&&\\
\ion{S}{2}  &  33\phm{Levels}  & 226\phm{Levels}  &    4166\phm{a}  & 13 \\*
\ion{S}{3}  &  41\phm{Levels}  & 235\phm{Levels}  &    3452\phm{a}  & 12 \\*
\ion{S}{4}  &  38\phm{Levels}  & 100\phm{Levels}  &     909\phm{a}  & 11 \\*
\ion{S}{5}  &  25\phm{Levels}  & 131\phm{Levels}  &    1171\phm{a}  & 10 \\*
\ion{S}{6}  &   1\phm{Levels}  &   1\phm{Levels}  &     \nodata     &                \\
&&&&\\
\ion{Fe}{2} &  36\phm{Levels}  &10\,921\phm{Levels}  &1\,264\,969\phm{a}  & 14, 15 \\*
\ion{Fe}{3} &  50\phm{Levels}  &12\,660\phm{Levels}  &1\,604\,934\phm{a}  & 14, 16 \\*
\ion{Fe}{4} &  43\phm{Levels}  &13\,705\phm{Levels}  &1\,776\,984\phm{a}  & 14, 17 \\*
\ion{Fe}{5} &  42\phm{Levels}  &11\,986\phm{Levels}  &1\,008\,385\phm{a}  & 14, 18 \\*
\ion{Fe}{6} &   1\phm{Levels}  &   1\phm{Levels}  &     \nodata           &                \\
\enddata
\tablerefs{(1) {\tt http://physics.nist.gov/PhysRefData/ASD/index.html}; (2) Luo~\& Pradhan 1989;
(3) Fernley et~al. 1999; (4) Tully, Seaton, \& Berrington 1990; (5) Peach, Saraph, \&
    Seaton 1988; (6) V.~M. Burke \& D.~J. Lennon, to be published; (7) K. Butler \&
    C.~J. Zeippen, to be published; (8) Hibbert \& Scott 1994; (9) K.~T. Taylor,
    to be published; (10) Butler, Mendoza, \& Zeippen 1993; (11) Mendoza et~al. 1995;
    (12) Nahar \& Pradhan 1993; (13) K. Butler, C. Mendoza, \& C.~J. Zeippen, to be published;
    (14) Kurucz 1994; (15) Nahar 1997; (16) Nahar 1996; (17) Bautista \& Pradhan 1997;
    (18) Bautista 1996.}
\end{deluxetable}

\clearpage
\begin{deluxetable}{ll}
\tablecaption{Key to the models' chemical compositions.  \label{KeyTbl}}
\tablehead{ \colhead{Key} & \colhead{Metallicity}}
\startdata
 C & $2\times\sun$  \\
 G & $1\times\sun$ \\
 L & $1/2\times\sun$  \\
 S & $1/5\times\sun$  \\
 T & $1/10\times\sun$ \\
 Z & $0$
\enddata
\end{deluxetable}

\clearpage
\begin{deluxetable}{llrrrrrrcrrrrrr}
\tabletypesize{\scriptsize}
\tablewidth{0pt}
\tablecaption{Bolometric corrections as function
   of effective temperature, gravity, microturbulent velocity and metallicity
   (6 metallicities from 2 times solar to metal-free models).  \label{BCTbl}}
\tablehead{
\colhead{ } & \colhead{ }   & \multicolumn{13}{c}{BC [mag]} \\
\colhead{$V_{\rm turb}$} & \colhead{ } & \multicolumn{6}{c}{2 \kms} & &\multicolumn{6}{c}{10 \kms} \\
\colhead{$Z/Z_\sun$} & \colhead{ } &  \colhead{2.} & \colhead{1.} & \colhead{0.5} &
\colhead{0.2} & \colhead{0.1} & \colhead{0.} & &\colhead{2.} & \colhead{1.} & \colhead{0.5} &
\colhead{0.2} & \colhead{0.1} & \colhead{0.} \\  \cline{3-8} \cline{10-15} 
\colhead{\teff\  [K]} & \colhead{$\log g$} & \multicolumn{13}{c}{ }  \\ [-3mm]
}
\startdata
 15,000\ldots &  1.75 & $-$1.29 &$-$1.32 & $-$1.33 & $-$1.35 & $-$1.36 & $-$1.39 & & $-$1.26 & $-$1.29 & $-$1.32 & $-$1.34 & $-$1.36 & $-$1.40 \\*
              &  2.00 & $-$1.23 &$-$1.26 & $-$1.28 & $-$1.30 & $-$1.31 & $-$1.33 & & $-$1.19 & $-$1.23 & $-$1.26 & $-$1.29 & $-$1.30 & $-$1.34 \\*
              &  2.25 & $-$1.21 &$-$1.24 & $-$1.26 & $-$1.28 & $-$1.29 & $-$1.31 & & $-$1.17 & $-$1.21 & $-$1.24 & $-$1.27 & $-$1.28 & $-$1.32 \\*
              &  2.50 & $-$1.20 &$-$1.23 & $-$1.26 & $-$1.28 & $-$1.29 & $-$1.30 & & $-$1.16 & $-$1.20 & $-$1.23 & $-$1.26 & $-$1.28 & $-$1.31 \\*
              &  2.75 & $-$1.20 &$-$1.23 & $-$1.25 & $-$1.27 & $-$1.28 & $-$1.30 & & $-$1.16 & $-$1.20 & $-$1.23 & $-$1.26 & $-$1.27 & $-$1.30 \\*
              &  3.00 & $-$1.20 &$-$1.23 & $-$1.25 & $-$1.27 & $-$1.28 & $-$1.30 & & $-$1.15 & $-$1.20 & $-$1.23 & $-$1.26 & $-$1.27 & $-$1.30 \\*
              &  3.25 & $-$1.20 &$-$1.23 & $-$1.25 & $-$1.27 & $-$1.28 & $-$1.30 & & \nodata & \nodata & \nodata & \nodata & \nodata & \nodata \\*
              &  3.50 & $-$1.19 &$-$1.23 & $-$1.25 & $-$1.27 & $-$1.28 & $-$1.30 & & \nodata & \nodata & \nodata & \nodata & \nodata & \nodata \\*
              &  3.75 & $-$1.19 &$-$1.22 & $-$1.25 & $-$1.27 & $-$1.28 & $-$1.29 & & \nodata & \nodata & \nodata & \nodata & \nodata & \nodata \\*
              &  4.00 & $-$1.19 &$-$1.22 & $-$1.25 & $-$1.27 & $-$1.28 & $-$1.29 & & \nodata & \nodata & \nodata & \nodata & \nodata & \nodata \\*
              &  4.25 & $-$1.19 &$-$1.22 & $-$1.24 & $-$1.26 & $-$1.27 & $-$1.29 & & \nodata & \nodata & \nodata & \nodata & \nodata & \nodata \\*
              &  4.50 & $-$1.19 &$-$1.22 & $-$1.24 & $-$1.26 & $-$1.27 & $-$1.29 & & \nodata & \nodata & \nodata & \nodata & \nodata & \nodata \\*
              &  4.75 & $-$1.19 &$-$1.22 & $-$1.24 & $-$1.26 & $-$1.27 & $-$1.29 & & \nodata & \nodata & \nodata & \nodata & \nodata & \nodata \\
 16,000\ldots &  2.00 & $-$1.39 &$-$1.42 & $-$1.44 & $-$1.46 & $-$1.47 & $-$1.50 & & $-$1.34 & $-$1.38 & $-$1.41 & $-$1.44 & $-$1.46 & $-$1.50 \\*
              &  2.25 & $-$1.37 &$-$1.40 & $-$1.42 & $-$1.44 & $-$1.45 & $-$1.47 & & $-$1.32 & $-$1.36 & $-$1.39 & $-$1.42 & $-$1.44 & $-$1.48 \\*
              &  2.50 & $-$1.36 &$-$1.39 & $-$1.41 & $-$1.43 & $-$1.44 & $-$1.46 & & $-$1.31 & $-$1.35 & $-$1.39 & $-$1.42 & $-$1.43 & $-$1.47 \\*
              &  2.75 & $-$1.36 &$-$1.39 & $-$1.41 & $-$1.43 & $-$1.44 & $-$1.46 & & $-$1.31 & $-$1.35 & $-$1.39 & $-$1.42 & $-$1.43 & $-$1.46 \\*
              &  3.00 & $-$1.36 &$-$1.39 & $-$1.41 & $-$1.43 & $-$1.44 & $-$1.46 & & $-$1.31 & $-$1.36 & $-$1.39 & $-$1.42 & $-$1.43 & $-$1.46 \\*
              &  3.25 & $-$1.36 &$-$1.39 & $-$1.41 & $-$1.43 & $-$1.44 & $-$1.46 & & \nodata & \nodata & \nodata & \nodata & \nodata & \nodata \\*
              &  3.50 & $-$1.36 &$-$1.39 & $-$1.41 & $-$1.43 & $-$1.44 & $-$1.46 & & \nodata & \nodata & \nodata & \nodata & \nodata & \nodata \\*
              &  3.75 & $-$1.36 &$-$1.39 & $-$1.41 & $-$1.43 & $-$1.44 & $-$1.46 & & \nodata & \nodata & \nodata & \nodata & \nodata & \nodata \\*
              &  4.00 & $-$1.36 &$-$1.39 & $-$1.41 & $-$1.43 & $-$1.44 & $-$1.46 & & \nodata & \nodata & \nodata & \nodata & \nodata & \nodata \\*
              &  4.25 & $-$1.36 &$-$1.39 & $-$1.41 & $-$1.43 & $-$1.44 & $-$1.46 & & \nodata & \nodata & \nodata & \nodata & \nodata & \nodata \\*
              &  4.50 & $-$1.36 &$-$1.39 & $-$1.41 & $-$1.43 & $-$1.44 & $-$1.45 & & \nodata & \nodata & \nodata & \nodata & \nodata & \nodata \\*
              &  4.75 & $-$1.35 &$-$1.38 & $-$1.41 & $-$1.43 & $-$1.44 & $-$1.45 & & \nodata & \nodata & \nodata & \nodata & \nodata & \nodata \\
 17,000\ldots &  2.00 & $-$1.54 &$-$1.57 & $-$1.59 & $-$1.61 & $-$1.62 & $-$1.66 & & $-$1.51 & $-$1.53 & $-$1.56 & $-$1.59 & $-$1.61 & $-$1.67 \\*
              &  2.25 & $-$1.51 &$-$1.54 & $-$1.56 & $-$1.58 & $-$1.59 & $-$1.62 & & $-$1.45 & $-$1.49 & $-$1.53 & $-$1.56 & $-$1.58 & $-$1.63 \\*
              &  2.50 & $-$1.50 &$-$1.53 & $-$1.56 & $-$1.58 & $-$1.59 & $-$1.61 & & $-$1.45 & $-$1.49 & $-$1.52 & $-$1.56 & $-$1.57 & $-$1.61 \\*
              &  2.75 & $-$1.50 &$-$1.53 & $-$1.56 & $-$1.58 & $-$1.59 & $-$1.61 & & $-$1.45 & $-$1.50 & $-$1.53 & $-$1.56 & $-$1.57 & $-$1.61 \\*
              &  3.00 & $-$1.51 &$-$1.54 & $-$1.56 & $-$1.58 & $-$1.59 & $-$1.61 & & $-$1.46 & $-$1.50 & $-$1.53 & $-$1.56 & $-$1.58 & $-$1.61 \\*
              &  3.25 & $-$1.51 &$-$1.54 & $-$1.56 & $-$1.58 & $-$1.59 & $-$1.61 & & \nodata & \nodata & \nodata & \nodata & \nodata & \nodata \\*
              &  3.50 & $-$1.51 &$-$1.54 & $-$1.57 & $-$1.58 & $-$1.59 & $-$1.61 & & \nodata & \nodata & \nodata & \nodata & \nodata & \nodata \\*
              &  3.75 & $-$1.51 &$-$1.54 & $-$1.57 & $-$1.59 & $-$1.59 & $-$1.61 & & \nodata & \nodata & \nodata & \nodata & \nodata & \nodata \\*
              &  4.00 & $-$1.52 &$-$1.55 & $-$1.57 & $-$1.59 & $-$1.60 & $-$1.61 & & \nodata & \nodata & \nodata & \nodata & \nodata & \nodata \\*
              &  4.25 & $-$1.52 &$-$1.55 & $-$1.57 & $-$1.59 & $-$1.60 & $-$1.61 & & \nodata & \nodata & \nodata & \nodata & \nodata & \nodata \\*
              &  4.50 & $-$1.52 &$-$1.55 & $-$1.57 & $-$1.59 & $-$1.59 & $-$1.61 & & \nodata & \nodata & \nodata & \nodata & \nodata & \nodata \\*
              &  4.75 & $-$1.51 &$-$1.54 & $-$1.57 & $-$1.58 & $-$1.59 & $-$1.61 & & \nodata & \nodata & \nodata & \nodata & \nodata & \nodata \\
 18,000\ldots &  2.00 & $-$1.72 &$-$1.74 & $-$1.75 & $-$1.77 & $-$1.79 & $-$1.83 & & \nodata & \nodata & $-$1.76 & $-$1.78 & $-$1.80 & $-$1.85 \\*
              &  2.25 & $-$1.64 &$-$1.67 & $-$1.69 & $-$1.72 & $-$1.73 & $-$1.77 & & $-$1.58 & $-$1.62 & $-$1.65 & $-$1.69 & $-$1.71 & $-$1.77 \\*
              &  2.50 & $-$1.63 &$-$1.67 & $-$1.69 & $-$1.71 & $-$1.72 & $-$1.75 & & $-$1.57 & $-$1.61 & $-$1.65 & $-$1.69 & $-$1.70 & $-$1.76 \\*
              &  2.75 & $-$1.64 &$-$1.67 & $-$1.69 & $-$1.71 & $-$1.72 & $-$1.75 & & $-$1.58 & $-$1.62 & $-$1.66 & $-$1.69 & $-$1.71 & $-$1.75 \\*
              &  3.00 & $-$1.64 &$-$1.68 & $-$1.70 & $-$1.72 & $-$1.73 & $-$1.75 & & $-$1.59 & $-$1.63 & $-$1.67 & $-$1.70 & $-$1.71 & $-$1.75 \\*
              &  3.25 & $-$1.65 &$-$1.68 & $-$1.70 & $-$1.72 & $-$1.73 & $-$1.75 & & \nodata & \nodata & \nodata & \nodata & \nodata & \nodata \\*
              &  3.50 & $-$1.65 &$-$1.69 & $-$1.71 & $-$1.73 & $-$1.74 & $-$1.75 & & \nodata & \nodata & \nodata & \nodata & \nodata & \nodata \\*
              &  3.75 & $-$1.66 &$-$1.69 & $-$1.71 & $-$1.73 & $-$1.74 & $-$1.76 & & \nodata & \nodata & \nodata & \nodata & \nodata & \nodata \\*
              &  4.00 & $-$1.66 &$-$1.69 & $-$1.71 & $-$1.73 & $-$1.74 & $-$1.76 & & \nodata & \nodata & \nodata & \nodata & \nodata & \nodata \\*
              &  4.25 & $-$1.66 &$-$1.69 & $-$1.72 & $-$1.73 & $-$1.74 & $-$1.76 & & \nodata & \nodata & \nodata & \nodata & \nodata & \nodata \\*
              &  4.50 & $-$1.66 &$-$1.69 & $-$1.72 & $-$1.73 & $-$1.74 & $-$1.76 & & \nodata & \nodata & \nodata & \nodata & \nodata & \nodata \\*
              &  4.75 & $-$1.66 &$-$1.69 & $-$1.72 & $-$1.73 & $-$1.74 & $-$1.75 & & \nodata & \nodata & \nodata & \nodata & \nodata & \nodata \\
 19,000\ldots &  2.25 & $-$1.77 &$-$1.79 & $-$1.82 & $-$1.84 & $-$1.86 & $-$1.90 & & $-$1.73 & $-$1.76 & $-$1.79 & $-$1.82 & $-$1.84 & $-$1.91 \\*
              &  2.50 & $-$1.75 &$-$1.79 & $-$1.81 & $-$1.84 & $-$1.85 & $-$1.89 & & $-$1.69 & $-$1.73 & $-$1.77 & $-$1.81 & $-$1.83 & $-$1.89 \\*
              &  2.75 & $-$1.76 &$-$1.79 & $-$1.82 & $-$1.84 & $-$1.85 & $-$1.88 & & $-$1.69 & $-$1.74 & $-$1.78 & $-$1.81 & $-$1.83 & $-$1.88 \\*
              &  3.00 & $-$1.77 &$-$1.80 & $-$1.83 & $-$1.85 & $-$1.86 & $-$1.88 & & $-$1.71 & $-$1.76 & $-$1.79 & $-$1.82 & $-$1.84 & $-$1.88 \\*
              &  3.25 & $-$1.78 &$-$1.81 & $-$1.83 & $-$1.85 & $-$1.86 & $-$1.89 & & \nodata & \nodata & \nodata & \nodata & \nodata & \nodata \\*
              &  3.50 & $-$1.79 &$-$1.82 & $-$1.84 & $-$1.86 & $-$1.87 & $-$1.89 & & \nodata & \nodata & \nodata & \nodata & \nodata & \nodata \\*
              &  3.75 & $-$1.79 &$-$1.82 & $-$1.85 & $-$1.86 & $-$1.87 & $-$1.89 & & \nodata & \nodata & \nodata & \nodata & \nodata & \nodata \\*
              &  4.00 & $-$1.80 &$-$1.83 & $-$1.85 & $-$1.87 & $-$1.88 & $-$1.89 & & \nodata & \nodata & \nodata & \nodata & \nodata & \nodata \\*
              &  4.25 & $-$1.80 &$-$1.83 & $-$1.85 & $-$1.87 & $-$1.88 & $-$1.89 & & \nodata & \nodata & \nodata & \nodata & \nodata & \nodata \\*
              &  4.50 & $-$1.80 &$-$1.83 & $-$1.85 & $-$1.87 & $-$1.88 & $-$1.89 & & \nodata & \nodata & \nodata & \nodata & \nodata & \nodata \\*
              &  4.75 & $-$1.81 &$-$1.83 & $-$1.86 & $-$1.87 & $-$1.88 & $-$1.89 & & \nodata & \nodata & \nodata & \nodata & \nodata & \nodata \\
 20,000\ldots &  2.25 & $-$1.91 &$-$1.93 & $-$1.95 & $-$1.98 & $-$1.99 & $-$2.04 & & $-$1.91 & $-$1.92 & $-$1.94 & $-$1.97 & $-$1.99 & $-$2.05 \\*
              &  2.50 & $-$1.87 &$-$1.90 & $-$1.93 & $-$1.95 & $-$1.97 & $-$2.01 & & $-$1.81 & $-$1.85 & $-$1.88 & $-$1.92 & $-$1.95 & $-$2.01 \\*
              &  2.75 & $-$1.87 &$-$1.91 & $-$1.93 & $-$1.96 & $-$1.97 & $-$2.01 & & $-$1.80 & $-$1.85 & $-$1.89 & $-$1.93 & $-$1.95 & $-$2.01 \\*
              &  3.00 & $-$1.89 &$-$1.92 & $-$1.95 & $-$1.97 & $-$1.98 & $-$2.01 & & $-$1.82 & $-$1.87 & $-$1.90 & $-$1.94 & $-$1.96 & $-$2.01 \\*
              &  3.25 & $-$1.90 &$-$1.93 & $-$1.96 & $-$1.98 & $-$1.99 & $-$2.01 & & \nodata & \nodata & \nodata & \nodata & \nodata & \nodata \\*
              &  3.50 & $-$1.91 &$-$1.94 & $-$1.96 & $-$1.98 & $-$1.99 & $-$2.02 & & \nodata & \nodata & \nodata & \nodata & \nodata & \nodata \\*
              &  3.75 & $-$1.92 &$-$1.95 & $-$1.97 & $-$1.99 & $-$2.00 & $-$2.02 & & \nodata & \nodata & \nodata & \nodata & \nodata & \nodata \\*
              &  4.00 & $-$1.92 &$-$1.96 & $-$1.97 & $-$2.00 & $-$2.00 & $-$2.02 & & \nodata & \nodata & \nodata & \nodata & \nodata & \nodata \\*
              &  4.25 & $-$1.93 &$-$1.96 & $-$1.98 & $-$2.00 & $-$2.01 & $-$2.02 & & \nodata & \nodata & \nodata & \nodata & \nodata & \nodata \\*
              &  4.50 & $-$1.93 &$-$1.96 & $-$1.99 & $-$2.00 & $-$2.01 & $-$2.03 & & \nodata & \nodata & \nodata & \nodata & \nodata & \nodata \\*
              &  4.75 & $-$1.94 &$-$1.97 & $-$1.99 & $-$2.00 & $-$2.01 & $-$2.02 & & \nodata & \nodata & \nodata & \nodata & \nodata & \nodata \\
 21,000\ldots &  2.25 & \nodata &\nodata & \nodata & $-$2.16 & $-$2.17 & $-$2.22 & & \nodata & \nodata & \nodata & $-$2.16 & $-$2.18 & $-$2.23 \\*
              &  2.50 & $-$1.98 &$-$2.01 & $-$2.04 & $-$2.06 & $-$2.08 & $-$2.12 & & $-$1.94 & $-$1.97 & $-$2.01 & $-$2.04 & $-$2.07 & $-$2.13 \\*
              &  2.75 & $-$1.98 &$-$2.01 & $-$2.04 & $-$2.07 & $-$2.09 & $-$2.13 & & $-$1.92 & $-$1.96 & $-$2.00 & $-$2.04 & $-$2.06 & $-$2.13 \\*
              &  3.00 & $-$1.99 &$-$2.03 & $-$2.06 & $-$2.08 & $-$2.10 & $-$2.13 & & $-$1.92 & $-$1.97 & $-$2.01 & $-$2.05 & $-$2.07 & $-$2.13 \\*
              &  3.25 & $-$2.01 &$-$2.04 & $-$2.07 & $-$2.09 & $-$2.10 & $-$2.14 & & \nodata & \nodata & \nodata & \nodata & \nodata & \nodata \\*
              &  3.50 & $-$2.02 &$-$2.06 & $-$2.08 & $-$2.10 & $-$2.11 & $-$2.14 & & \nodata & \nodata & \nodata & \nodata & \nodata & \nodata \\*
              &  3.75 & $-$2.03 &$-$2.07 & $-$2.09 & $-$2.11 & $-$2.12 & $-$2.14 & & \nodata & \nodata & \nodata & \nodata & \nodata & \nodata \\*
              &  4.00 & $-$2.04 &$-$2.08 & $-$2.09 & $-$2.12 & $-$2.13 & $-$2.15 & & \nodata & \nodata & \nodata & \nodata & \nodata & \nodata \\*
              &  4.25 & $-$2.05 &$-$2.08 & $-$2.10 & $-$2.12 & $-$2.13 & $-$2.15 & & \nodata & \nodata & \nodata & \nodata & \nodata & \nodata \\*
              &  4.50 & $-$2.06 &$-$2.09 & $-$2.11 & $-$2.13 & $-$2.13 & $-$2.15 & & \nodata & \nodata & \nodata & \nodata & \nodata & \nodata \\*
              &  4.75 & $-$2.06 &$-$2.09 & $-$2.11 & $-$2.13 & $-$2.13 & $-$2.15 & & \nodata & \nodata & \nodata & \nodata & \nodata & \nodata \\
 22,000\ldots &  2.50 & $-$2.10 &$-$2.12 & $-$2.15 & $-$2.18 & $-$2.20 & $-$2.24 & & $-$2.07 & $-$2.10 & $-$2.13 & $-$2.17 & $-$2.19 & $-$2.24 \\*
              &  2.75 & $-$2.08 &$-$2.12 & $-$2.15 & $-$2.17 & $-$2.19 & $-$2.24 & & $-$2.03 & $-$2.07 & $-$2.11 & $-$2.15 & $-$2.17 & $-$2.24 \\*
              &  3.00 & $-$2.10 &$-$2.13 & $-$2.16 & $-$2.19 & $-$2.20 & $-$2.24 & & $-$2.03 & $-$2.07 & $-$2.11 & $-$2.15 & $-$2.18 & $-$2.24 \\*
              &  3.25 & $-$2.11 &$-$2.15 & $-$2.18 & $-$2.20 & $-$2.21 & $-$2.25 & & \nodata & \nodata & \nodata & \nodata & \nodata & \nodata \\*
              &  3.50 & $-$2.13 &$-$2.16 & $-$2.19 & $-$2.21 & $-$2.22 & $-$2.26 & & \nodata & \nodata & \nodata & \nodata & \nodata & \nodata \\*
              &  3.75 & $-$2.14 &$-$2.18 & $-$2.20 & $-$2.22 & $-$2.23 & $-$2.26 & & \nodata & \nodata & \nodata & \nodata & \nodata & \nodata \\*
              &  4.00 & $-$2.15 &$-$2.19 & $-$2.21 & $-$2.23 & $-$2.24 & $-$2.26 & & \nodata & \nodata & \nodata & \nodata & \nodata & \nodata \\*
              &  4.25 & $-$2.16 &$-$2.19 & $-$2.22 & $-$2.24 & $-$2.25 & $-$2.27 & & \nodata & \nodata & \nodata & \nodata & \nodata & \nodata \\*
              &  4.50 & $-$2.17 &$-$2.20 & $-$2.22 & $-$2.24 & $-$2.25 & $-$2.27 & & \nodata & \nodata & \nodata & \nodata & \nodata & \nodata \\*
              &  4.75 & $-$2.18 &$-$2.21 & $-$2.23 & $-$2.24 & $-$2.25 & $-$2.27 & & \nodata & \nodata & \nodata & \nodata & \nodata & \nodata \\
 23,000\ldots &  2.50 & $-$2.24 &$-$2.26 & $-$2.28 & $-$2.31 & $-$2.32 & $-$2.37 & & $-$2.25 & $-$2.25 & $-$2.27 & $-$2.30 & $-$2.32 & $-$2.38 \\*
              &  2.75 & $-$2.19 &$-$2.22 & $-$2.25 & $-$2.28 & $-$2.29 & $-$2.33 & & $-$2.14 & $-$2.18 & $-$2.22 & $-$2.26 & $-$2.28 & $-$2.33 \\*
              &  3.00 & $-$2.20 &$-$2.23 & $-$2.26 & $-$2.29 & $-$2.30 & $-$2.35 & & $-$2.14 & $-$2.18 & $-$2.22 & $-$2.26 & $-$2.28 & $-$2.35 \\*
              &  3.25 & $-$2.21 &$-$2.25 & $-$2.27 & $-$2.30 & $-$2.32 & $-$2.36 & & \nodata & \nodata & \nodata & \nodata & \nodata & \nodata \\*
              &  3.50 & $-$2.23 &$-$2.26 & $-$2.29 & $-$2.32 & $-$2.33 & $-$2.36 & & \nodata & \nodata & \nodata & \nodata & \nodata & \nodata \\*
              &  3.75 & $-$2.24 &$-$2.28 & $-$2.30 & $-$2.33 & $-$2.34 & $-$2.37 & & \nodata & \nodata & \nodata & \nodata & \nodata & \nodata \\*
              &  4.00 & $-$2.26 &$-$2.29 & $-$2.32 & $-$2.34 & $-$2.35 & $-$2.38 & & \nodata & \nodata & \nodata & \nodata & \nodata & \nodata \\*
              &  4.25 & $-$2.27 &$-$2.30 & $-$2.32 & $-$2.34 & $-$2.35 & $-$2.38 & & \nodata & \nodata & \nodata & \nodata & \nodata & \nodata \\*
              &  4.50 & $-$2.28 &$-$2.31 & $-$2.33 & $-$2.35 & $-$2.36 & $-$2.38 & & \nodata & \nodata & \nodata & \nodata & \nodata & \nodata \\*
              &  4.75 & $-$2.28 &$-$2.32 & $-$2.34 & $-$2.35 & $-$2.36 & $-$2.38 & & \nodata & \nodata & \nodata & \nodata & \nodata & \nodata \\
 24,000\ldots &  2.50 & $-$2.43 &$-$2.43 & $-$2.43 & $-$2.45 & $-$2.47 & $-$2.52 & & $-$2.49 & $-$2.46 & $-$2.45 & $-$2.46 & $-$2.47 & $-$2.53 \\*
              &  2.75 & $-$2.29 &$-$2.32 & $-$2.35 & $-$2.37 & $-$2.39 & $-$2.43 & & $-$2.25 & $-$2.29 & $-$2.32 & $-$2.36 & $-$2.38 & $-$2.44 \\*
              &  3.00 & $-$2.29 &$-$2.33 & $-$2.35 & $-$2.38 & $-$2.40 & $-$2.44 & & $-$2.24 & $-$2.28 & $-$2.32 & $-$2.36 & $-$2.39 & $-$2.44 \\*
              &  3.25 & $-$2.31 &$-$2.34 & $-$2.37 & $-$2.40 & $-$2.41 & $-$2.46 & & \nodata & \nodata & \nodata & \nodata & \nodata & \nodata \\*
              &  3.50 & $-$2.32 &$-$2.36 & $-$2.39 & $-$2.41 & $-$2.43 & $-$2.47 & & \nodata & \nodata & \nodata & \nodata & \nodata & \nodata \\*
              &  3.75 & $-$2.34 &$-$2.37 & $-$2.40 & $-$2.43 & $-$2.44 & $-$2.47 & & \nodata & \nodata & \nodata & \nodata & \nodata & \nodata \\*
              &  4.00 & $-$2.35 &$-$2.39 & $-$2.41 & $-$2.44 & $-$2.45 & $-$2.48 & & \nodata & \nodata & \nodata & \nodata & \nodata & \nodata \\*
              &  4.25 & $-$2.37 &$-$2.40 & $-$2.42 & $-$2.45 & $-$2.46 & $-$2.48 & & \nodata & \nodata & \nodata & \nodata & \nodata & \nodata \\*
              &  4.50 & $-$2.38 &$-$2.41 & $-$2.43 & $-$2.45 & $-$2.46 & $-$2.49 & & \nodata & \nodata & \nodata & \nodata & \nodata & \nodata \\*
              &  4.75 & $-$2.39 &$-$2.42 & $-$2.44 & $-$2.46 & $-$2.47 & $-$2.49 & & \nodata & \nodata & \nodata & \nodata & \nodata & \nodata \\
 25,000\ldots &  2.75 & $-$2.40 &$-$2.42 & $-$2.45 & $-$2.48 & $-$2.50 & $-$2.54 & & $-$2.38 & $-$2.40 & $-$2.43 & $-$2.46 & $-$2.49 & $-$2.55 \\*
              &  3.00 & $-$2.38 &$-$2.42 & $-$2.45 & $-$2.47 & $-$2.49 & $-$2.53 & & $-$2.34 & $-$2.38 & $-$2.42 & $-$2.46 & $-$2.48 & $-$2.53 \\*
              &  3.25 & $-$2.40 &$-$2.43 & $-$2.46 & $-$2.49 & $-$2.51 & $-$2.55 & & \nodata & \nodata & \nodata & \nodata & \nodata & \nodata \\*
              &  3.50 & $-$2.42 &$-$2.45 & $-$2.48 & $-$2.51 & $-$2.52 & $-$2.56 & & \nodata & \nodata & \nodata & \nodata & \nodata & \nodata \\*
              &  3.75 & $-$2.43 &$-$2.47 & $-$2.49 & $-$2.52 & $-$2.53 & $-$2.57 & & \nodata & \nodata & \nodata & \nodata & \nodata & \nodata \\*
              &  4.00 & $-$2.45 &$-$2.48 & $-$2.51 & $-$2.53 & $-$2.55 & $-$2.58 & & \nodata & \nodata & \nodata & \nodata & \nodata & \nodata \\*
              &  4.25 & $-$2.46 &$-$2.49 & $-$2.52 & $-$2.54 & $-$2.55 & $-$2.58 & & \nodata & \nodata & \nodata & \nodata & \nodata & \nodata \\*
              &  4.50 & $-$2.47 &$-$2.50 & $-$2.53 & $-$2.55 & $-$2.56 & $-$2.59 & & \nodata & \nodata & \nodata & \nodata & \nodata & \nodata \\*
              &  4.75 & $-$2.48 &$-$2.51 & $-$2.54 & $-$2.56 & $-$2.57 & $-$2.59 & & \nodata & \nodata & \nodata & \nodata & \nodata & \nodata \\
 26,000\ldots &  2.75 & $-$2.52 &$-$2.54 & $-$2.56 & $-$2.58 & $-$2.60 & $-$2.65 & & $-$2.52 & $-$2.53 & $-$2.55 & $-$2.57 & $-$2.59 & $-$2.66 \\*
              &  3.00 & $-$2.48 &$-$2.51 & $-$2.53 & $-$2.56 & $-$2.58 & $-$2.62 & & $-$2.43 & $-$2.47 & $-$2.51 & $-$2.55 & $-$2.57 & $-$2.62 \\*
              &  3.25 & $-$2.49 &$-$2.52 & $-$2.55 & $-$2.58 & $-$2.59 & $-$2.63 & & \nodata & \nodata & \nodata & \nodata & \nodata & \nodata \\*
              &  3.50 & $-$2.51 &$-$2.54 & $-$2.57 & $-$2.60 & $-$2.61 & $-$2.65 & & \nodata & \nodata & \nodata & \nodata & \nodata & \nodata \\*
              &  3.75 & $-$2.52 &$-$2.56 & $-$2.58 & $-$2.61 & $-$2.62 & $-$2.66 & & \nodata & \nodata & \nodata & \nodata & \nodata & \nodata \\*
              &  4.00 & $-$2.54 &$-$2.57 & $-$2.60 & $-$2.62 & $-$2.64 & $-$2.67 & & \nodata & \nodata & \nodata & \nodata & \nodata & \nodata \\*
              &  4.25 & $-$2.55 &$-$2.58 & $-$2.61 & $-$2.63 & $-$2.65 & $-$2.68 & & \nodata & \nodata & \nodata & \nodata & \nodata & \nodata \\*
              &  4.50 & $-$2.56 &$-$2.59 & $-$2.62 & $-$2.64 & $-$2.65 & $-$2.68 & & \nodata & \nodata & \nodata & \nodata & \nodata & \nodata \\*
              &  4.75 & $-$2.57 &$-$2.60 & $-$2.63 & $-$2.65 & $-$2.66 & $-$2.69 & & \nodata & \nodata & \nodata & \nodata & \nodata & \nodata \\
 27,000\ldots &  2.75 & $-$2.68 &$-$2.68 & $-$2.69 & $-$2.71 & $-$2.72 & $-$2.77 & & $-$2.68 & $-$2.70 & $-$2.70 & $-$2.71 & $-$2.72 & $-$2.77 \\*
              &  3.00 & $-$2.57 &$-$2.60 & $-$2.62 & $-$2.65 & $-$2.67 & $-$2.71 & & $-$2.54 & $-$2.57 & $-$2.60 & $-$2.64 & $-$2.66 & $-$2.72 \\*
              &  3.25 & $-$2.57 &$-$2.61 & $-$2.63 & $-$2.66 & $-$2.68 & $-$2.71 & & \nodata & \nodata & \nodata & \nodata & \nodata & \nodata \\*
              &  3.50 & $-$2.59 &$-$2.62 & $-$2.65 & $-$2.68 & $-$2.69 & $-$2.73 & & \nodata & \nodata & \nodata & \nodata & \nodata & \nodata \\*
              &  3.75 & $-$2.61 &$-$2.64 & $-$2.67 & $-$2.70 & $-$2.71 & $-$2.74 & & \nodata & \nodata & \nodata & \nodata & \nodata & \nodata \\*
              &  4.00 & $-$2.62 &$-$2.66 & $-$2.68 & $-$2.71 & $-$2.72 & $-$2.76 & & \nodata & \nodata & \nodata & \nodata & \nodata & \nodata \\*
              &  4.25 & $-$2.63 &$-$2.67 & $-$2.69 & $-$2.72 & $-$2.73 & $-$2.77 & & \nodata & \nodata & \nodata & \nodata & \nodata & \nodata \\*
              &  4.50 & $-$2.65 &$-$2.68 & $-$2.70 & $-$2.73 & $-$2.74 & $-$2.77 & & \nodata & \nodata & \nodata & \nodata & \nodata & \nodata \\*
              &  4.75 & $-$2.66 &$-$2.69 & $-$2.71 & $-$2.74 & $-$2.75 & $-$2.78 & & \nodata & \nodata & \nodata & \nodata & \nodata & \nodata \\
 28,000\ldots &  2.75 & $-$2.86 &$-$2.85 & $-$2.85 & $-$2.86 & $-$2.86 & $-$2.89 & & \nodata & \nodata & \nodata & $-$2.87 & \nodata & $-$2.90 \\*
              &  3.00 & $-$2.67 &$-$2.69 & $-$2.72 & $-$2.74 & $-$2.76 & $-$2.81 & & $-$2.65 & $-$2.67 & $-$2.70 & $-$2.73 & $-$2.75 & $-$2.81 \\*
              &  3.25 & $-$2.66 &$-$2.69 & $-$2.72 & $-$2.74 & $-$2.76 & $-$2.79 & & \nodata & \nodata & \nodata & \nodata & \nodata & \nodata \\*
              &  3.50 & $-$2.67 &$-$2.71 & $-$2.73 & $-$2.76 & $-$2.77 & $-$2.80 & & \nodata & \nodata & \nodata & \nodata & \nodata & \nodata \\*
              &  3.75 & $-$2.69 &$-$2.72 & $-$2.75 & $-$2.78 & $-$2.79 & $-$2.82 & & \nodata & \nodata & \nodata & \nodata & \nodata & \nodata \\*
              &  4.00 & $-$2.71 &$-$2.74 & $-$2.76 & $-$2.79 & $-$2.80 & $-$2.83 & & \nodata & \nodata & \nodata & \nodata & \nodata & \nodata \\*
              &  4.25 & $-$2.72 &$-$2.75 & $-$2.78 & $-$2.80 & $-$2.81 & $-$2.85 & & \nodata & \nodata & \nodata & \nodata & \nodata & \nodata \\*
              &  4.50 & $-$2.73 &$-$2.76 & $-$2.79 & $-$2.81 & $-$2.82 & $-$2.85 & & \nodata & \nodata & \nodata & \nodata & \nodata & \nodata \\*
              &  4.75 & $-$2.74 &$-$2.77 & $-$2.80 & $-$2.82 & $-$2.83 & $-$2.86 & & \nodata & \nodata & \nodata & \nodata & \nodata & \nodata \\
 29,000\ldots &  3.00 & $-$2.78 &$-$2.80 & $-$2.82 & $-$2.84 & $-$2.85 & $-$2.90 & & $-$2.78 & $-$2.79 & $-$2.80 & $-$2.83 & $-$2.85 & $-$2.90 \\*
              &  3.25 & $-$2.74 &$-$2.77 & $-$2.80 & $-$2.82 & $-$2.84 & $-$2.88 & & \nodata & \nodata & \nodata & \nodata & \nodata & \nodata \\*
              &  3.50 & $-$2.75 &$-$2.78 & $-$2.81 & $-$2.84 & $-$2.85 & $-$2.88 & & \nodata & \nodata & \nodata & \nodata & \nodata & \nodata \\*
              &  3.75 & $-$2.77 &$-$2.80 & $-$2.83 & $-$2.85 & $-$2.86 & $-$2.89 & & \nodata & \nodata & \nodata & \nodata & \nodata & \nodata \\*
              &  4.00 & $-$2.78 &$-$2.82 & $-$2.84 & $-$2.87 & $-$2.88 & $-$2.91 & & \nodata & \nodata & \nodata & \nodata & \nodata & \nodata \\*
              &  4.25 & $-$2.80 &$-$2.83 & $-$2.86 & $-$2.88 & $-$2.89 & $-$2.92 & & \nodata & \nodata & \nodata & \nodata & \nodata & \nodata \\*
              &  4.50 & $-$2.81 &$-$2.84 & $-$2.87 & $-$2.89 & $-$2.90 & $-$2.93 & & \nodata & \nodata & \nodata & \nodata & \nodata & \nodata \\*
              &  4.75 & $-$2.82 &$-$2.85 & $-$2.87 & $-$2.90 & $-$2.91 & $-$2.94 & & \nodata & \nodata & \nodata & \nodata & \nodata & \nodata \\
 30,000\ldots &  3.00 & $-$2.91 &$-$2.92 & $-$2.93 & $-$2.95 & $-$2.96 & $-$2.99 & & $-$2.91 & $-$2.92 & $-$2.92 & $-$2.94 & $-$2.95 & $-$3.00 \\*
              &  3.25 & $-$2.84 &$-$2.86 & $-$2.88 & $-$2.91 & $-$2.92 & $-$2.96 & & \nodata & \nodata & \nodata & \nodata & \nodata & \nodata \\*
              &  3.50 & $-$2.83 &$-$2.86 & $-$2.89 & $-$2.91 & $-$2.93 & $-$2.96 & & \nodata & \nodata & \nodata & \nodata & \nodata & \nodata \\*
              &  3.75 & $-$2.85 &$-$2.88 & $-$2.90 & $-$2.93 & $-$2.94 & $-$2.97 & & \nodata & \nodata & \nodata & \nodata & \nodata & \nodata \\*
              &  4.00 & $-$2.86 &$-$2.89 & $-$2.92 & $-$2.94 & $-$2.95 & $-$2.98 & & \nodata & \nodata & \nodata & \nodata & \nodata & \nodata \\*
              &  4.25 & $-$2.87 &$-$2.91 & $-$2.93 & $-$2.95 & $-$2.96 & $-$2.99 & & \nodata & \nodata & \nodata & \nodata & \nodata & \nodata \\*
              &  4.50 & $-$2.89 &$-$2.92 & $-$2.94 & $-$2.96 & $-$2.97 & $-$3.00 & & \nodata & \nodata & \nodata & \nodata & \nodata & \nodata \\*
              &  4.75 & $-$2.90 &$-$2.93 & $-$2.95 & $-$2.97 & $-$2.98 & $-$3.01 & & \nodata & \nodata & \nodata & \nodata & \nodata & \nodata
\enddata
\end{deluxetable}

\clearpage
\begin{deluxetable}{lccc}
\tablewidth{0pt}
\tablecaption{Ionizing fluxes in the \ion{H}{1} Lyman continuum as function
   of effective temperature and gravity for solar composition model atmospheres.  \label{Q0Tbl}}
\tablehead{
\colhead{ } & \multicolumn{3}{c}{
$q_0 = \log N_{\rm LyC}$ [s$^{-1}$ cm$^{-2}$]} \\
\colhead{\teff\  [K]} & \colhead{$\log g=4$}   & \colhead{$\log g=3$} & \colhead{$\log g=2$}
}
\startdata
 55000 &  25.01 & \nodata & \nodata \\*
 50000 &  24.82 & \nodata & \nodata \\*
 45000 &  24.58 & \nodata & \nodata \\*
 40000 &  24.28 & \nodata & \nodata \\*
 35000 &  23.84 & \nodata & \nodata \\*
 30000 &  22.90 &  23.62  & \nodata \\*
 29000 &  22.63 &  23.45  & \nodata \\*
 28000 &  22.35 &  23.25  & \nodata \\*
 27000 &  22.06 &  22.99  & \nodata \\*
 26000 &  21.76 &  22.66  & \nodata \\*
 25000 &  21.44 &  22.25  & \nodata \\*
 24000 &  21.12 &  21.82  & \nodata \\*
 23000 &  20.81 &  21.40  & \nodata \\*
 22000 &  20.51 &  20.95  & \nodata \\*
 21000 &  20.21 &  20.54  & \nodata \\*
 20000 &  19.90 &  20.16  & \nodata \\*
 19000 &  19.59 &  19.82  & \nodata \\*
 18000 &  19.26 &  19.47  &  20.74  \\*
 17000 &  18.92 &  19.11  &  19.63  \\*
 16000 &  18.57 &  18.74  &  19.11  \\*
 15000 &  18.20 &  18.35  &  18.65  \\*
\enddata
\end{deluxetable}

\clearpage
\begin{deluxetable}{lcccc}
\tablewidth{0pt}
\tablecaption{Eddington limit and effective gravities as function
   of effective temperature for solar composition model atmospheres.  \label{EffGravTbl}}
\tablehead{
\colhead{ } & \colhead{ } & \multicolumn{3}{c}{
$\log g_{\rm eff}$ [cgs]} \\ \cline{3-5}
\colhead{\teff\  [K]} & \colhead{$\log g_{\rm Edd}$} &
\colhead{$\log g=4$}   & \colhead{$\log g=3$} & \colhead{$\log g=2$}
}
\startdata
 55000 &  3.83  &  3.10 & \nodata & \nodata \\*
 50000 &  3.67  &  3.37 & \nodata & \nodata \\*
 45000 &  3.51  &  3.51 & \nodata & \nodata \\*
 40000 &  3.33  &  3.63 & \nodata & \nodata \\*
 35000 &  3.11  &  3.73 & \nodata & \nodata \\*
 30000 &  2.85  &  3.81 &  2.02   & \nodata \\*
 28000 &  2.73  &  3.83 &  2.22   & \nodata \\*
 26000 &  2.60  &  3.86 &  2.35   & \nodata \\*
 24000 &  2.46  &  3.88 &  2.48   & \nodata \\*
 22000 &  2.31  &  3.90 &  2.59   & \nodata \\*
 20000 &  2.14  &  3.92 &  2.69   & \nodata \\*
 18000 &  1.96  &  3.93 &  2.76   &  0.57   \\*
 15000 &  1.67  &  3.96 &  2.85   &  1.32   \\*
\enddata
\end{deluxetable}

\clearpage

\clearpage

\clearpage

\begin{figure}
\includegraphics[scale=0.79,angle=0]{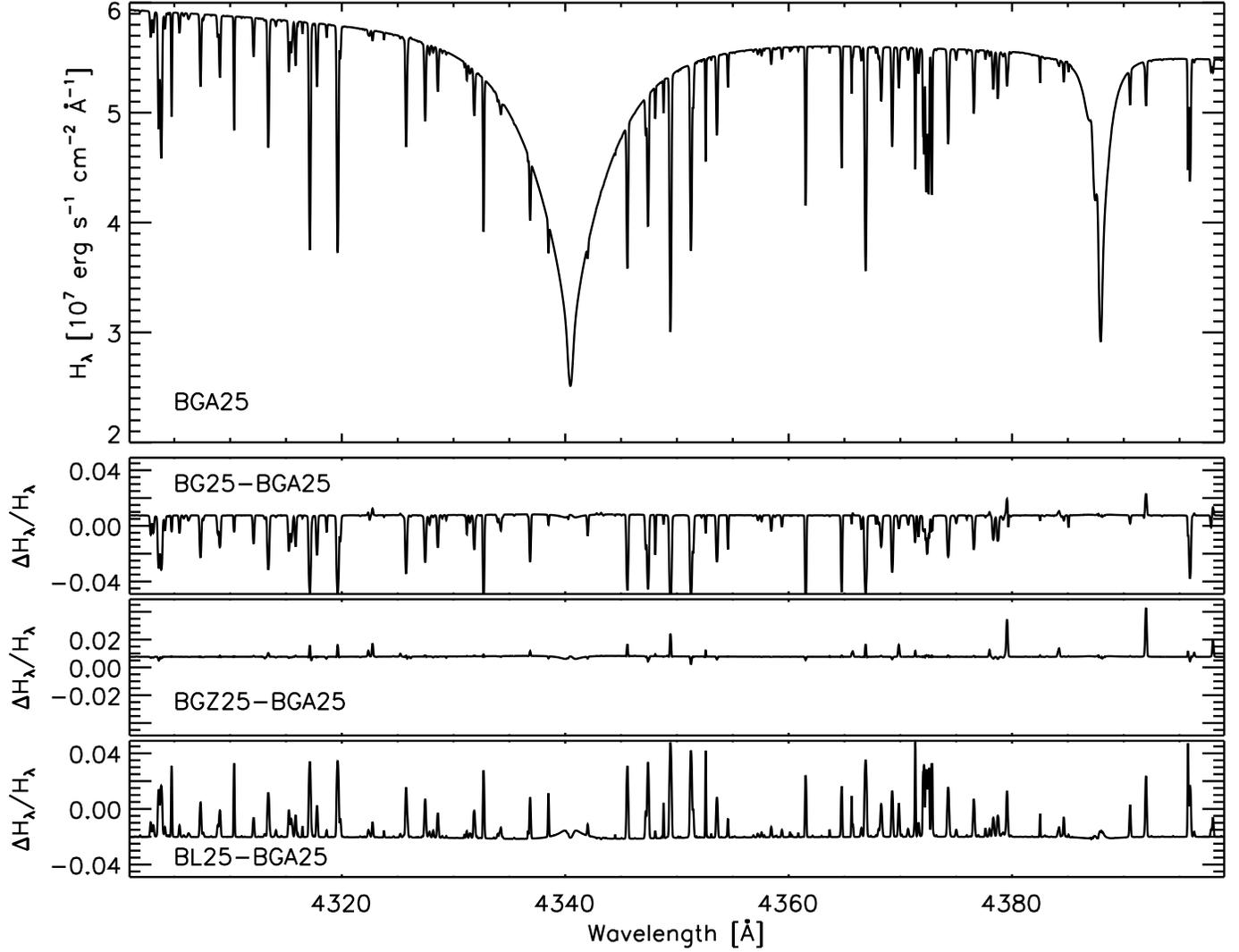}
\figurenum{3}
\caption{Model spectrum around H$\gamma$ for a model atmosphere with updated C, N,
O, and Ne abundances (\teff~=~25\,000\,K, $\log g=4.0$, solar composition
with new CNO and Ne abundances, and \vtur~=~2\,\kms). The bottom panels show
the relative spectrum changes on the same scale for models with various abundances
(see text).
\label{ABFig}}
\end{figure}

\clearpage

\begin{figure}
\includegraphics[scale=0.67,angle=0]{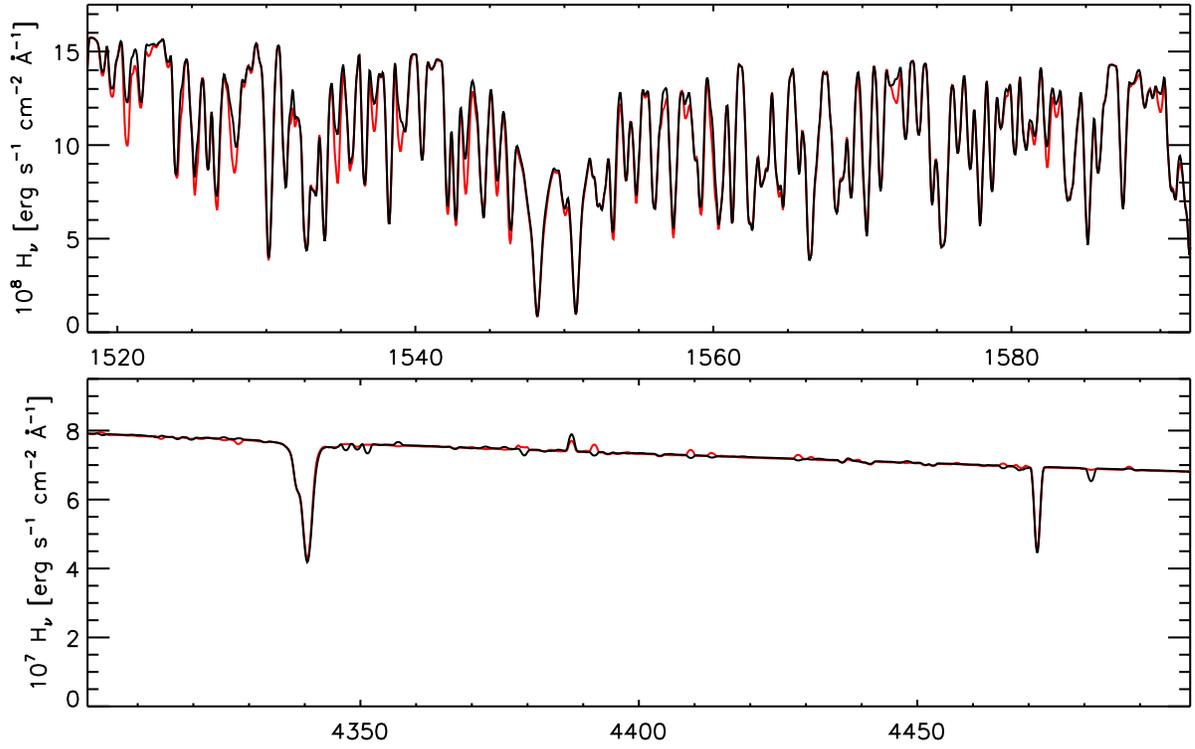}
\figurenum{4}
\caption{Ultraviolet and visible model spectra for model atmospheres with identical stellar
parameters (\teff~=~30\,000\,K, $\log g=3.0$, solar composition, and \vtur~=~10\,\kms) from
the two grids: \grid\  (black line) and {\sc Ostar2002} (grey line, red line in electronic
edition). A rotational broadening of 50\,\kms\  was applied to the spectra.
\label{O2B6Fig}}
\end{figure}

\clearpage

\begin{figure}
\includegraphics[scale=0.7,angle=0]{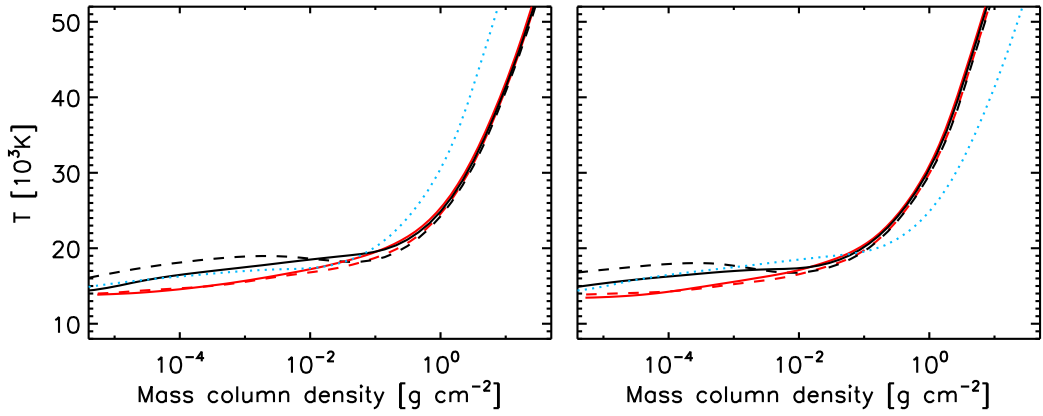}
\figurenum{5}
\caption{Temperature structure of model atmospheres with \teff = 25\,000\,K,
$\log g=3.0$ (left panel) and  $\log g=4.0$ (right panel). Black lines show the
temperature stratification of the NLTE BSTAR2006 models, compared to the
LTE Kurucz models (grey lines; red in electronic edition).
Solar composition models (full lines) and metal-poor
(1/10 solar) models (dashed lines) are displayed.
The dotted lines illustrate the effect of the different surface gravities.
\label{TtauFig}}
\end{figure}

\clearpage

\begin{figure}
\epsscale{0.9} \plotone{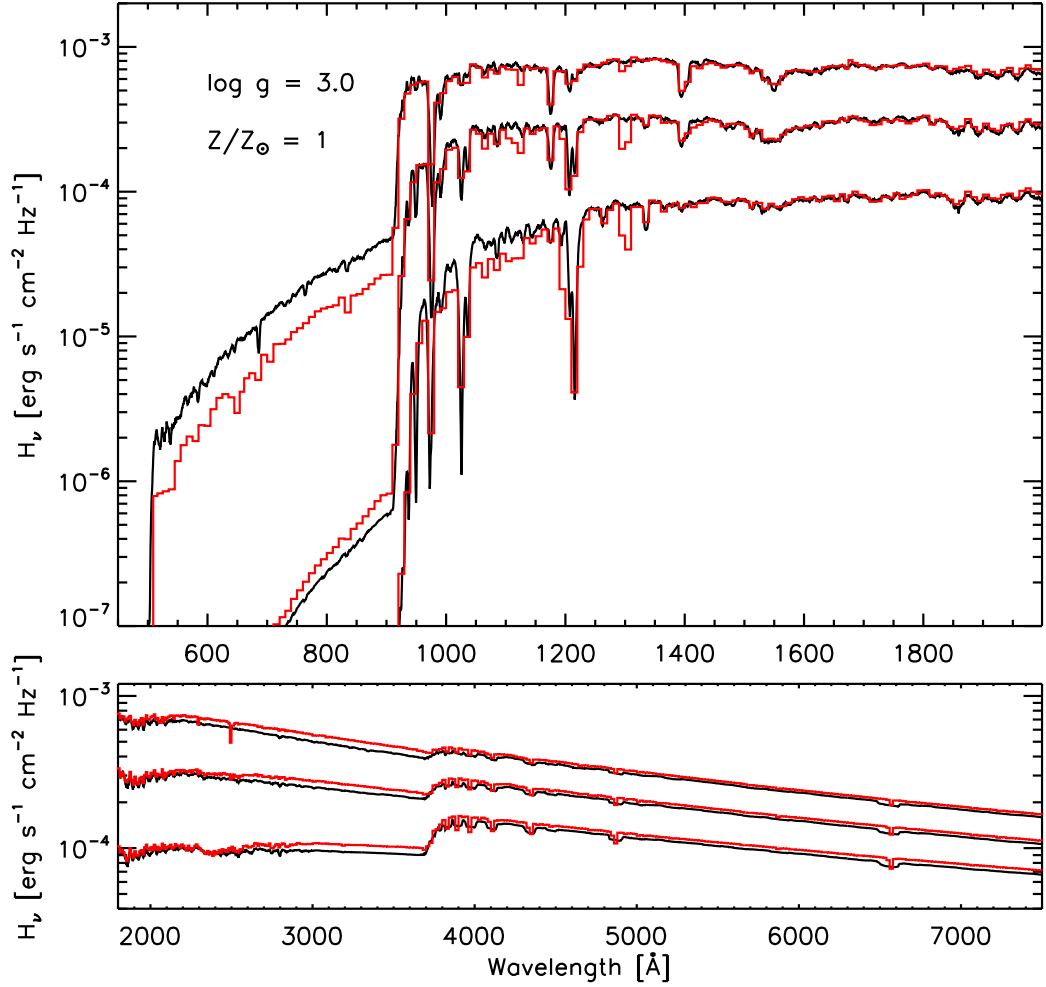}
\figurenum{6}
\caption{Predicted flux for three solar composition model atmospheres with $(T_{\rm eff},
     \log g)$ equal to (25000\,K, 3.0); (20000\,K, 3.0), and (15000\,K, 3.0) -- black lines;
     compared to Kurucz (1993) models with the same parameters -- grey histograms
     (red in electronic edition).  \label{FKurFig}}
\end{figure}

\clearpage

\begin{figure}
\epsscale{0.9} \plotone{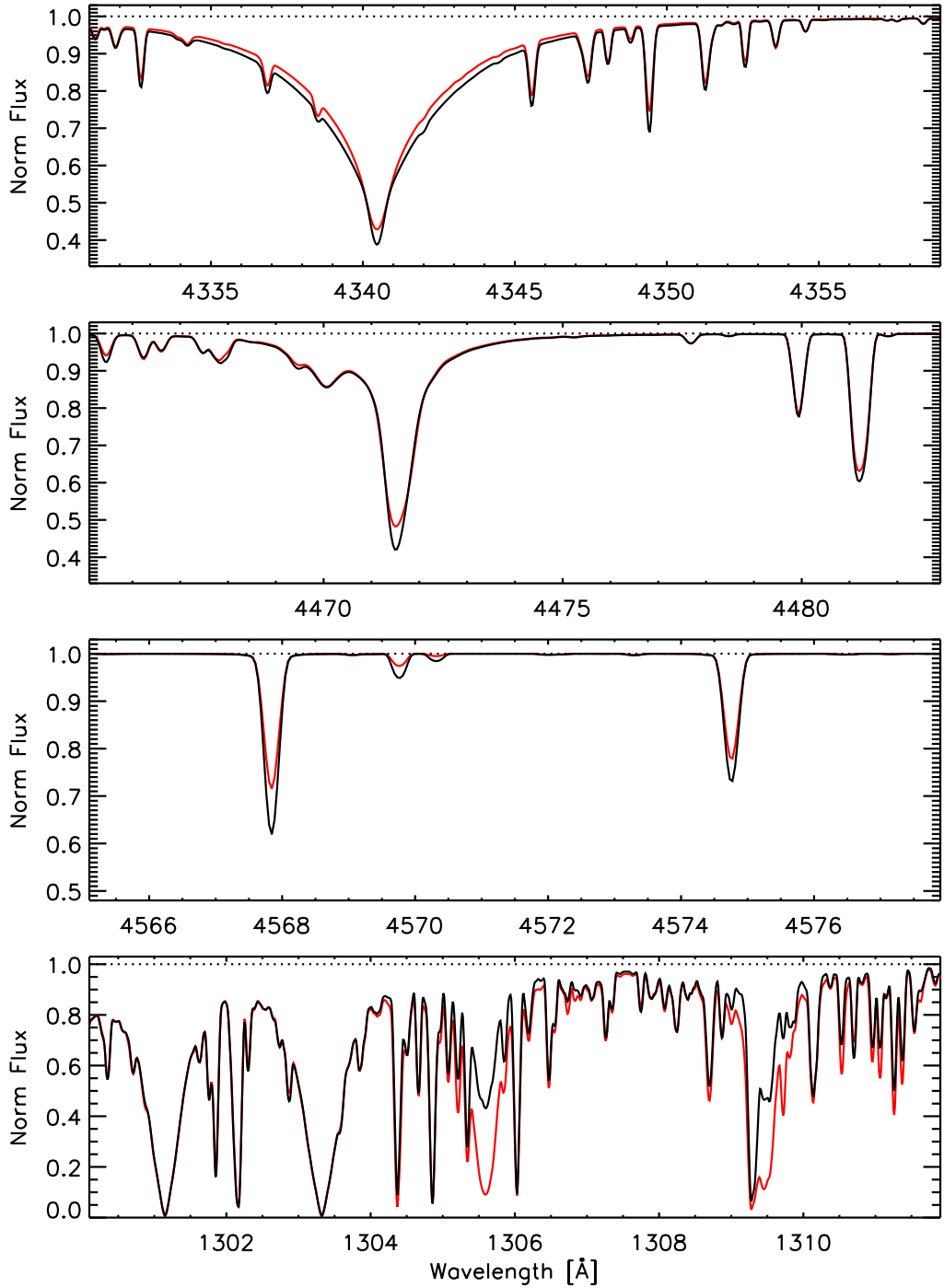}
\figurenum{7}
\caption{Predicted line profiles for a solar composition model atmospheres with 
     \teff~=~20\,000\,K, $\log g=3.0$, and \vtur~=~2\,\kms\ -- black lines;
     compared to the Kurucz (1993) model with the same parameters -- grey lines
     (red lines in electronic edition).  \label{LinesFig}}
\end{figure}

\clearpage

\begin{figure}
\includegraphics[scale=0.7,angle=90]{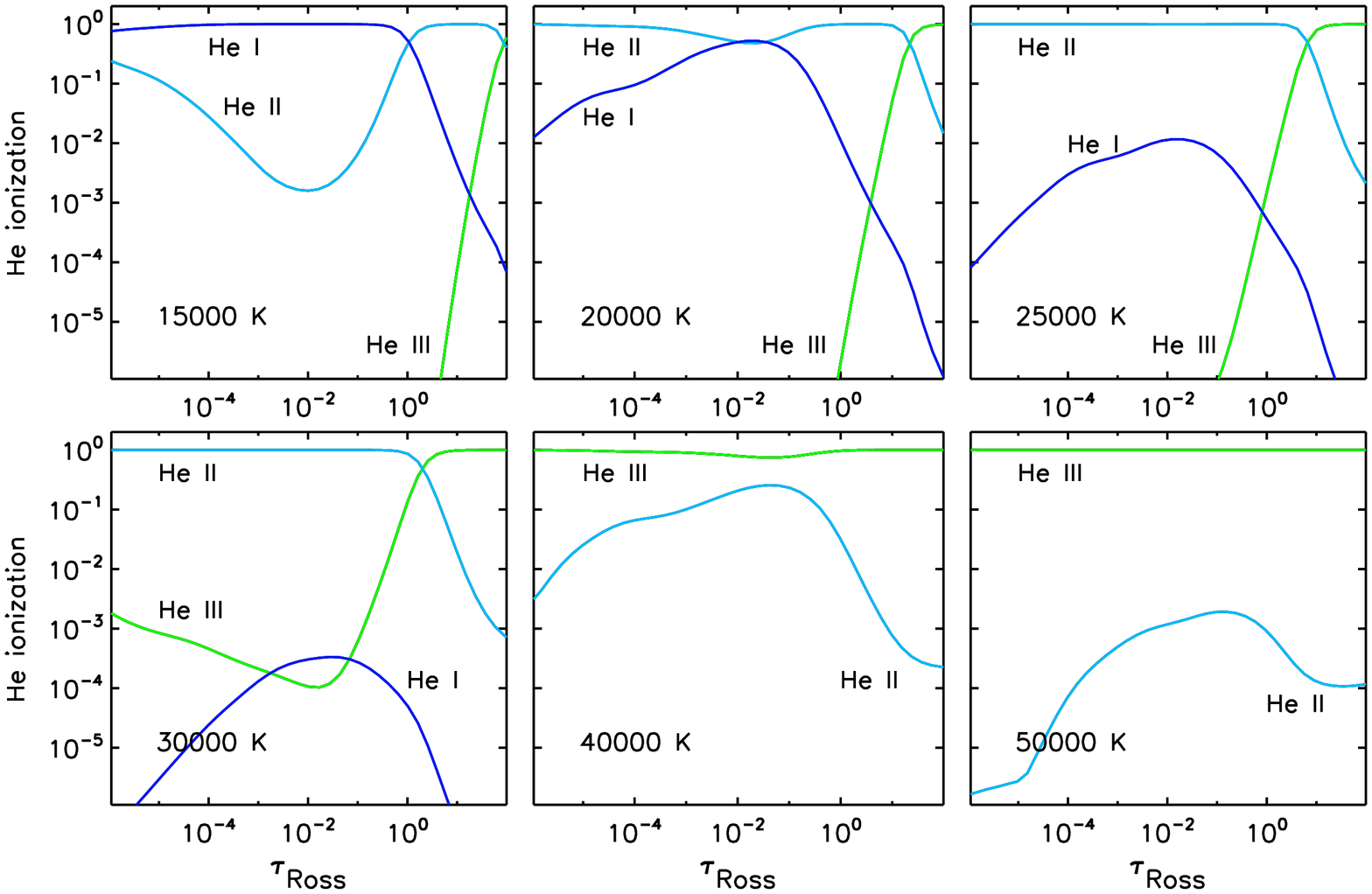}
\figurenum{8}
\caption{Ionization fractions of helium as function of depth in 6 NLTE model atmospheres, 
   \teff = 15\,000, 20\,000, 25\,000, 30\,000, 40\,000 and 50\,000\,K, 
   $\log g=4.0$, and solar composition.  \label{IonHeFig}}
\end{figure}

\clearpage

\begin{figure}
\includegraphics[scale=0.7,angle=90]{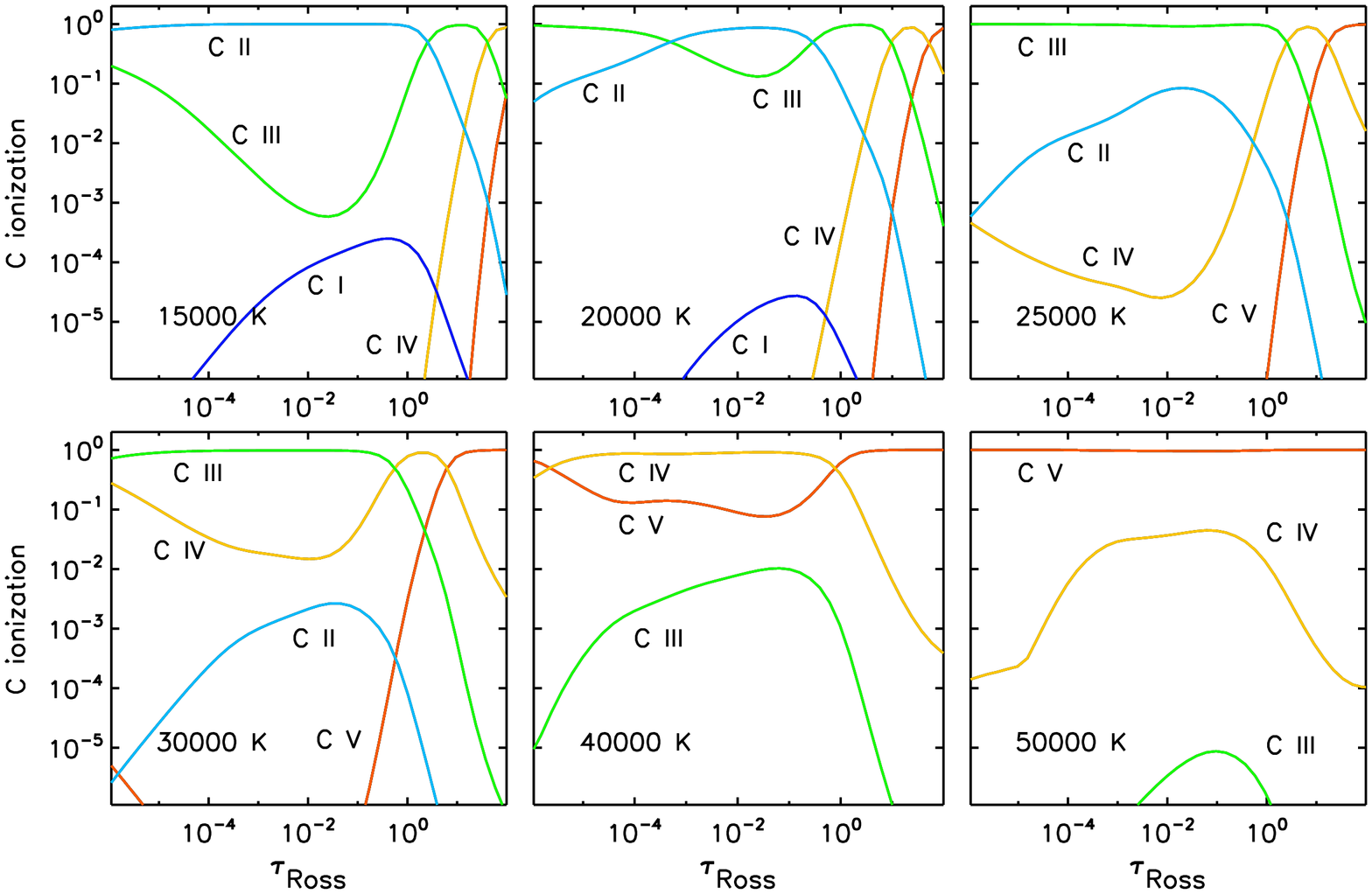}
\figurenum{9}
\caption{Ionization fractions of carbon as function of depth in 6 NLTE model atmospheres, 
   \teff = 15\,000, 20\,000, 25\,000, 30\,000, 40\,000 and 50\,000\,K, 
   $\log g=4.0$, and solar composition.  \label{IonCFig}}
\end{figure}

\clearpage

\begin{figure}
\includegraphics[scale=0.7,angle=90]{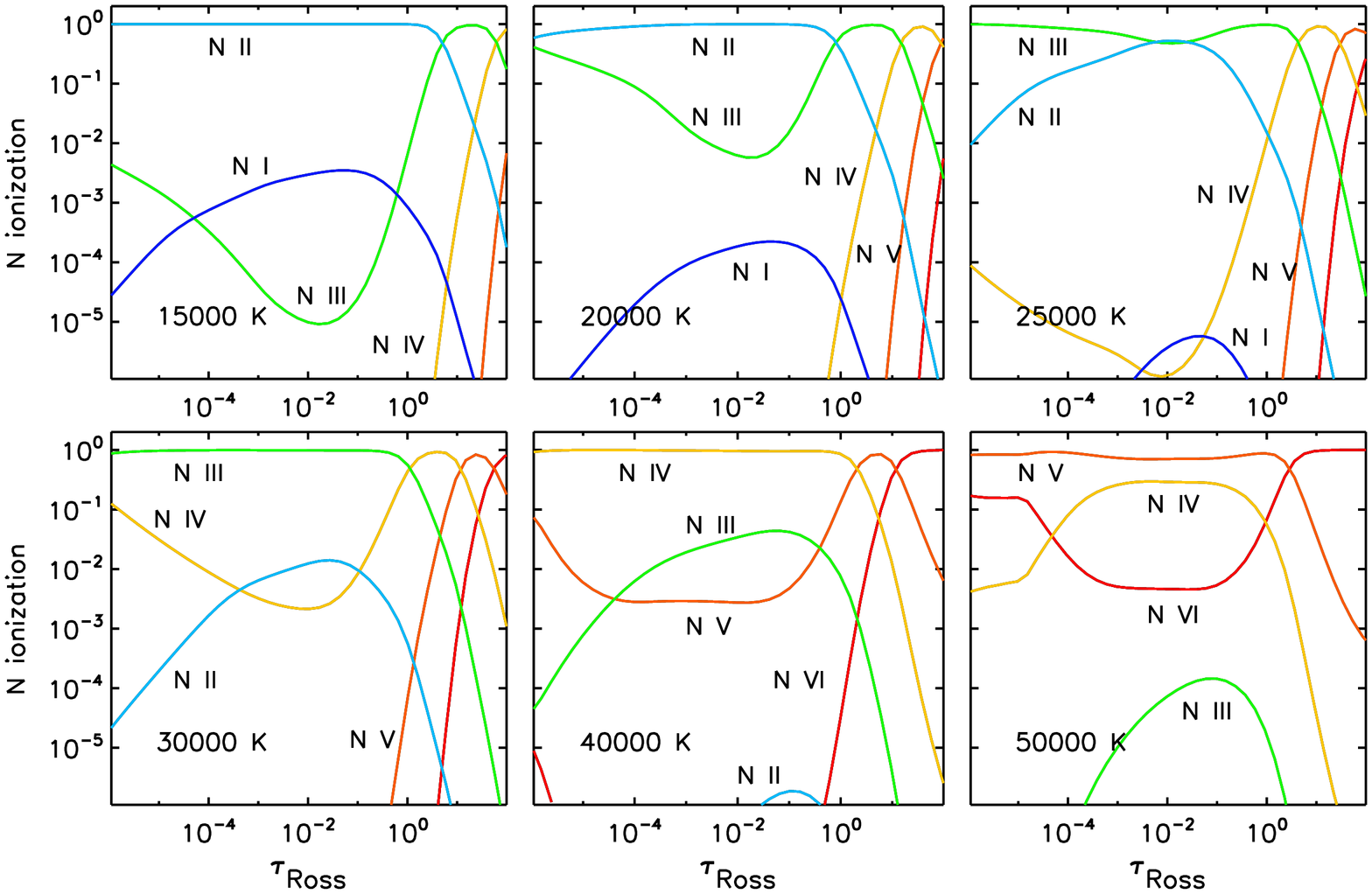}
\figurenum{10}
\caption{Ionization fractions of nitrogen as function of depth in 6 NLTE model atmospheres, 
   \teff = 15\,000, 20\,000, 25\,000, 30\,000, 40\,000 and 50\,000\,K, 
   $\log g=4.0$, and solar composition.  \label{IonNFig}}
\end{figure}

\clearpage

\begin{figure}
\includegraphics[scale=0.7,angle=90]{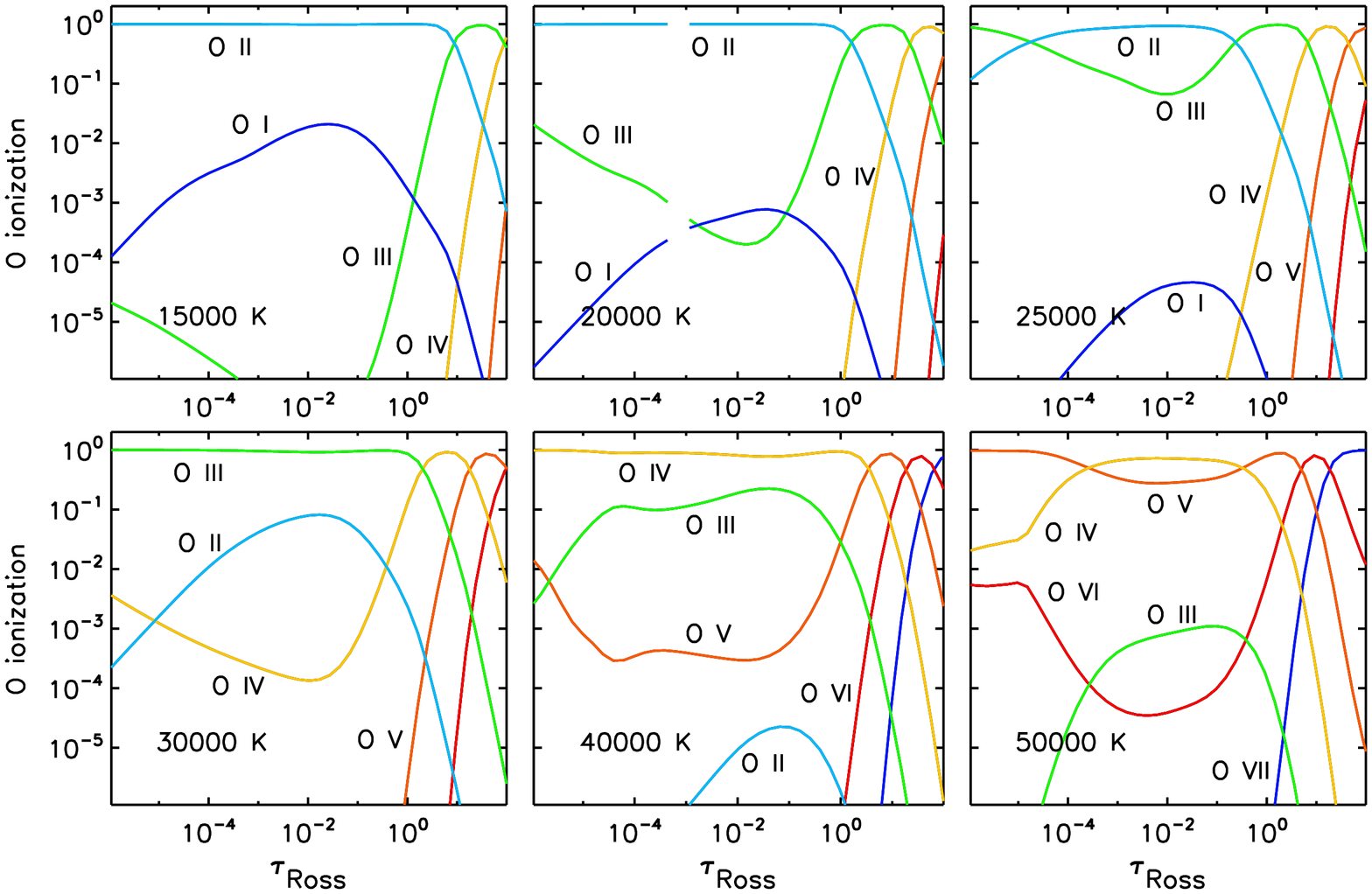}
\figurenum{11}
\caption{Ionization fractions of oxygen as function of depth in 6 NLTE model atmospheres, 
   \teff = 15\,000, 20\,000, 25\,000, 30\,000, 40\,000 and 50\,000\,K, 
   $\log g=4.0$, and solar composition.  \label{IonOFig}}
\end{figure}

\clearpage

\begin{figure}
\includegraphics[scale=0.7,angle=90]{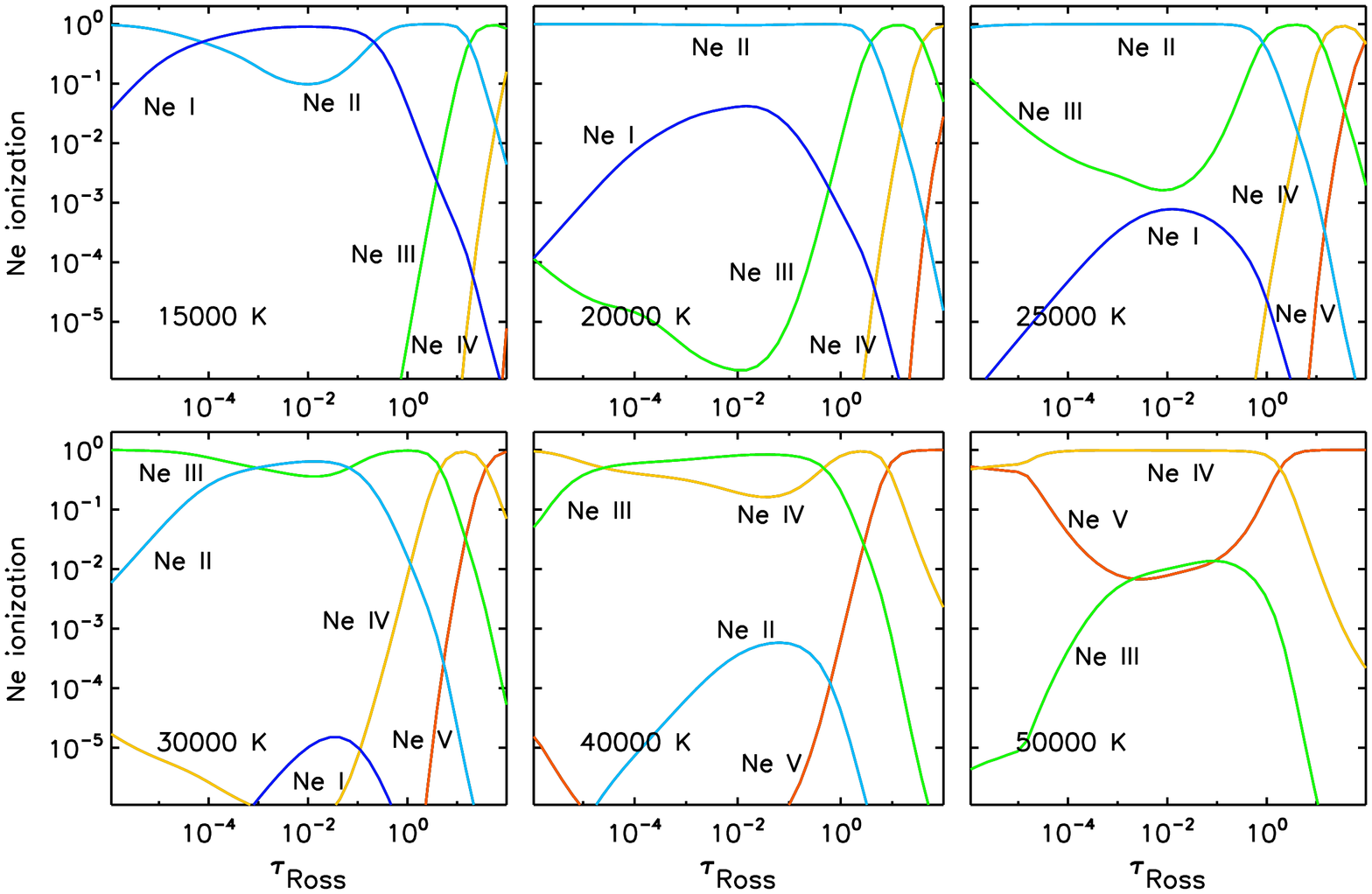}
\figurenum{12}
\caption{Ionization fractions of neon as function of depth in 6 NLTE model atmospheres, 
   \teff = 15\,000, 20\,000, 25\,000, 30\,000, 40\,000 and 50\,000\,K, 
   $\log g=4.0$, and solar composition.  \label{IonNeFig}}
\end{figure}

\clearpage

\begin{figure}
\includegraphics[scale=0.7,angle=90]{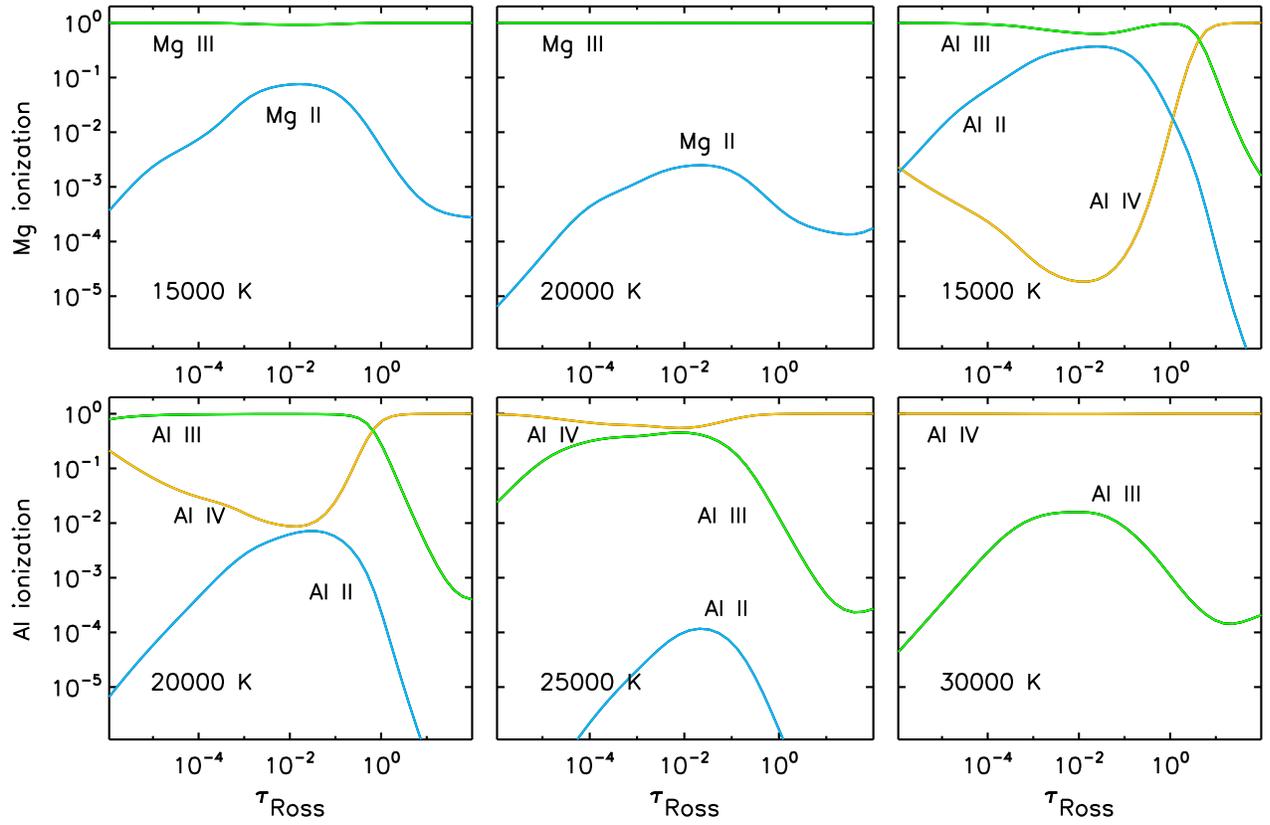}
\figurenum{13}
\caption{Ionization fractions of magnesium and aluminum as function of depth in 4 NLTE model atmospheres, 
   \teff = 15\,000, 20\,000, 25\,000, and 30\,000\,K, 
   $\log g=4.0$, and solar composition.  \label{IonMgAlFig}}
\end{figure}

\clearpage

\begin{figure}
\includegraphics[scale=0.7,angle=90]{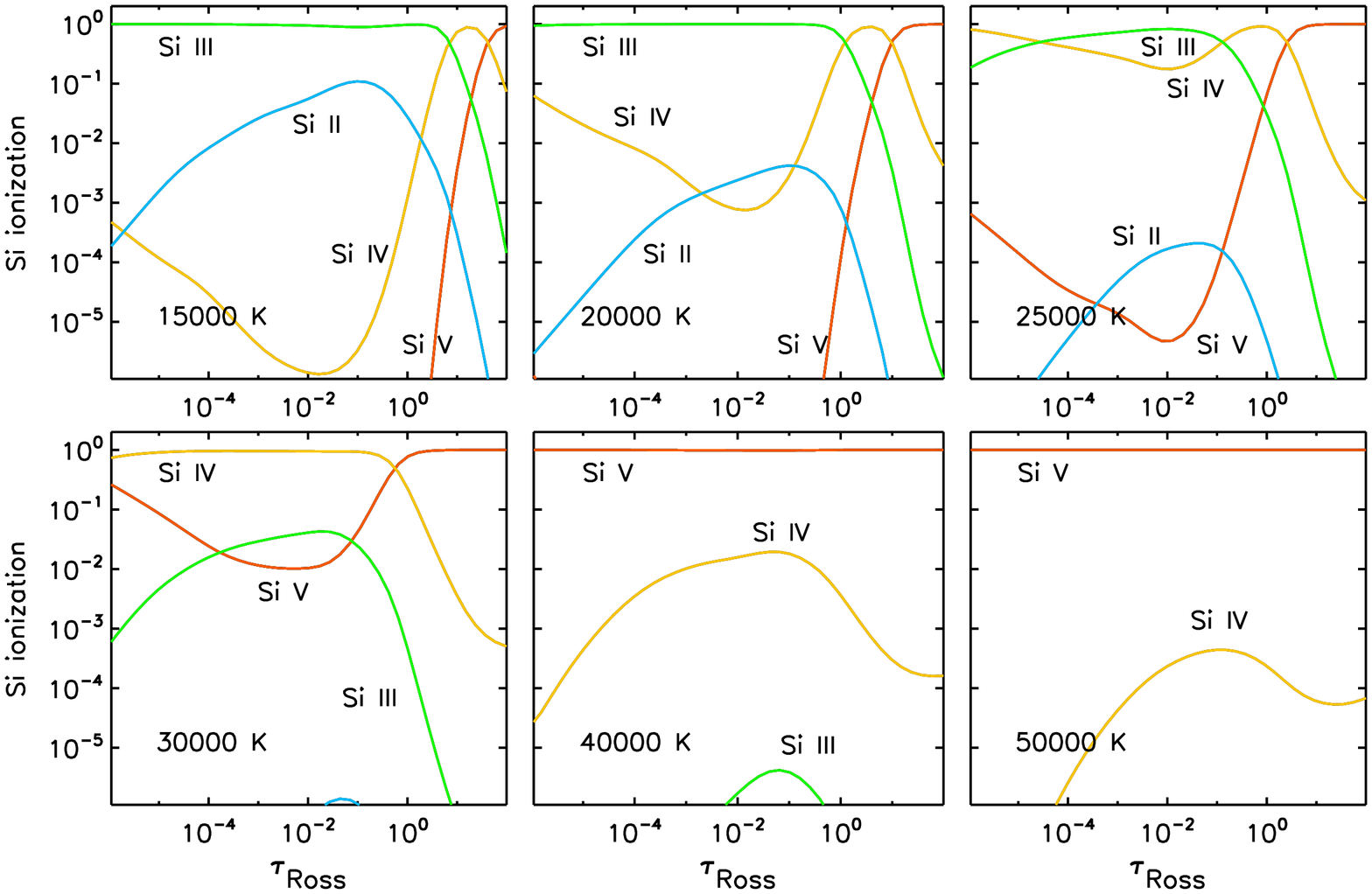}
\figurenum{14}
\caption{Ionization fractions of silicon as function of depth in 6 NLTE model atmospheres, 
   \teff = 15\,000, 20\,000, 25\,000, 30\,000, 40\,000 and 50\,000\,K, 
   $\log g=4.0$, and solar composition.  \label{IonSiFig}}
\end{figure}

\clearpage

\begin{figure}
\includegraphics[scale=0.7,angle=90]{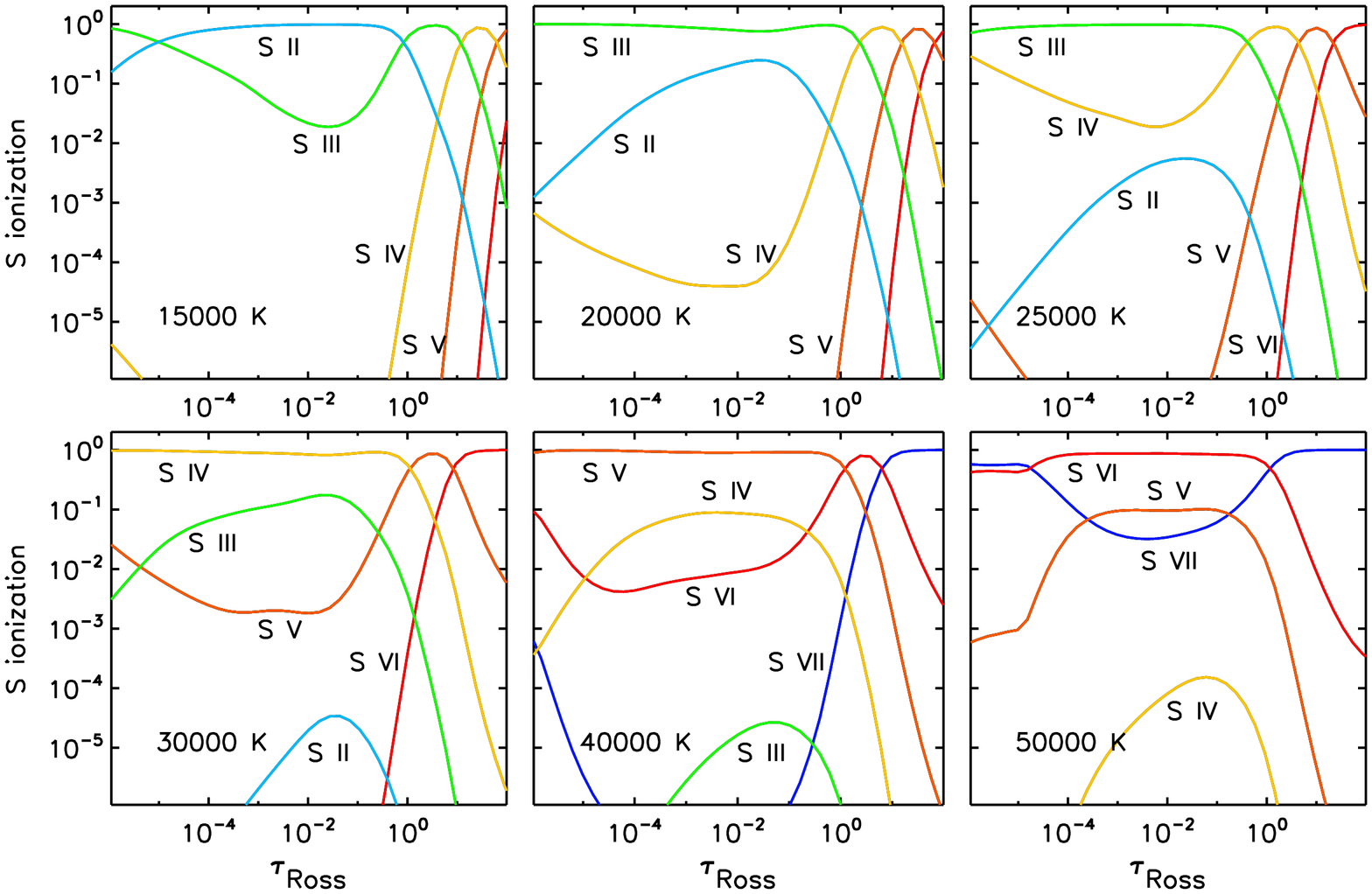}
\figurenum{15}
\caption{Ionization fractions of sulfur as function of depth in 6 NLTE model atmospheres, 
   \teff = 15\,000, 20\,000, 25\,000, 30\,000, 40\,000 and 50\,000\,K, 
   $\log g=4.0$, and solar composition.  \label{IonSFig}}
\end{figure}

\clearpage

\begin{figure}
\includegraphics[scale=0.7,angle=90]{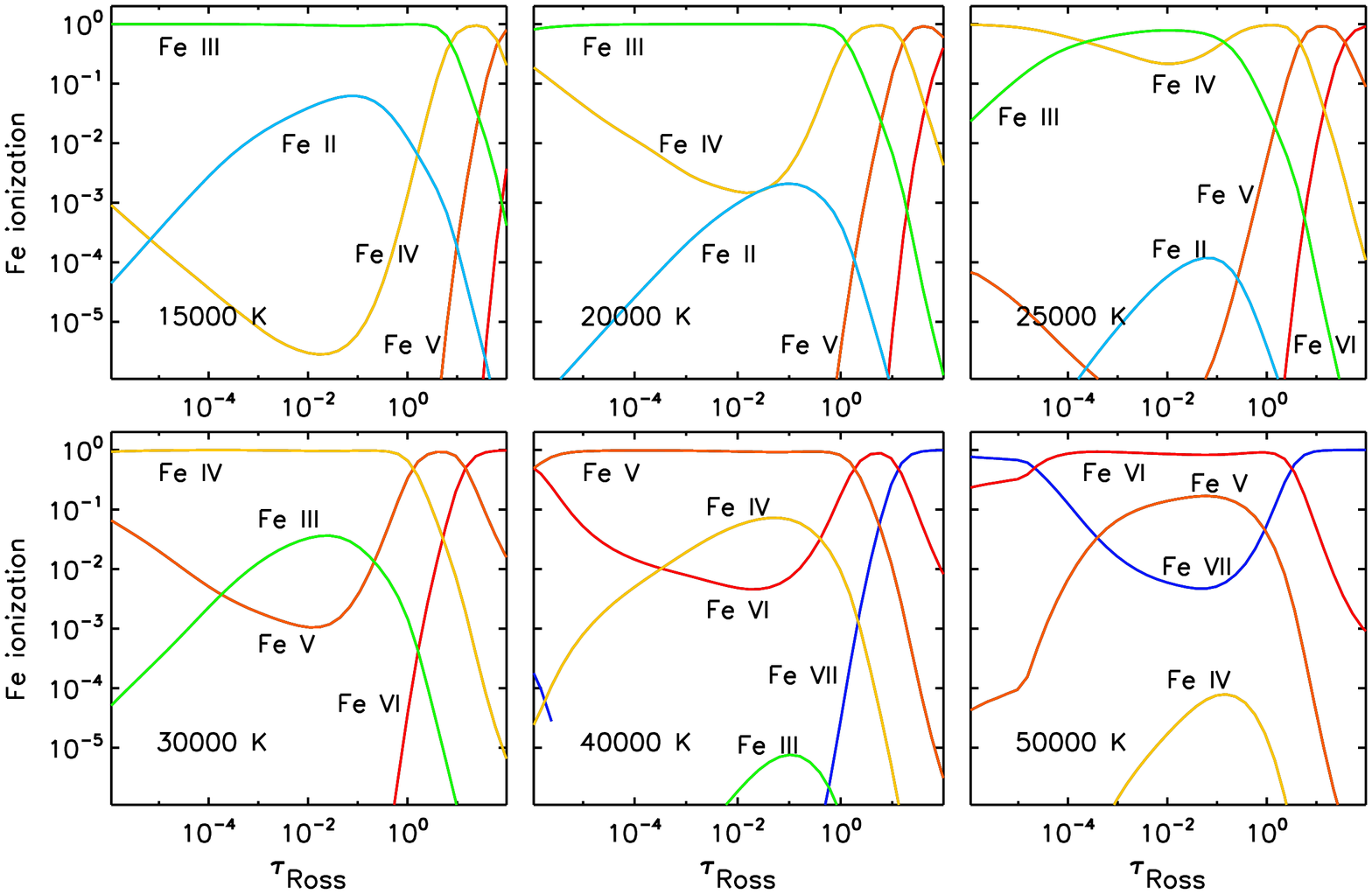}
\figurenum{16}
\caption{Ionization fractions of iron as function of depth in 6 NLTE model atmospheres, 
   \teff = 15\,000, 20\,000, 25\,000, 30\,000, 40\,000 and 50\,000\,K, 
   $\log g=4.0$, and solar composition.  \label{IonFeFig}}
\end{figure}

\clearpage

\begin{figure}
\includegraphics[scale=0.7,angle=90]{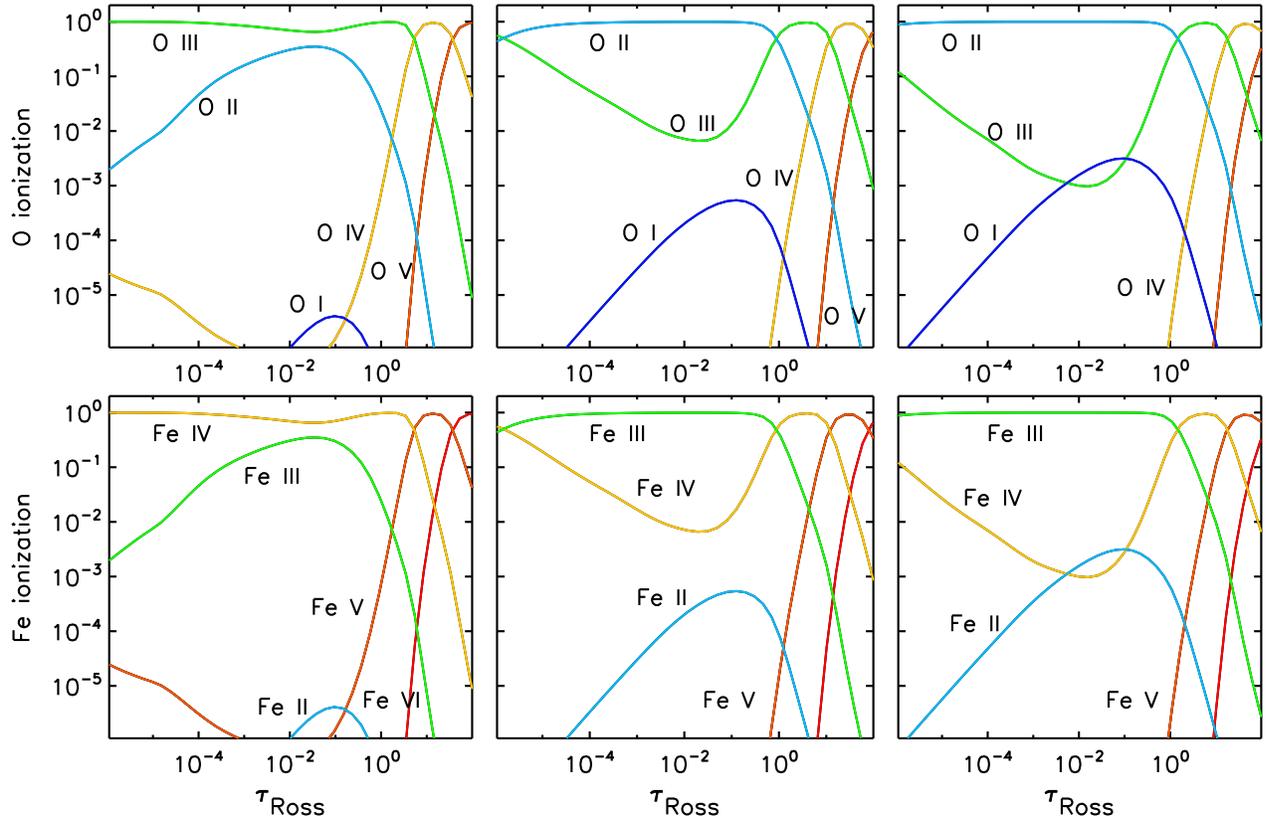}
\figurenum{17}
\caption{Ionization fractions of oxygen and iron as function of depth in 3 NLTE model atmospheres
   with different surface gravities ($\log g = 2.25, 3.25, 4.25$, from left to right). Solar
   composition and \teff = 20\,000\,K is assumed for the 3 models. \label{IonGravFig}}
\end{figure}

\clearpage

\begin{figure}
\epsscale{0.9} \plotone{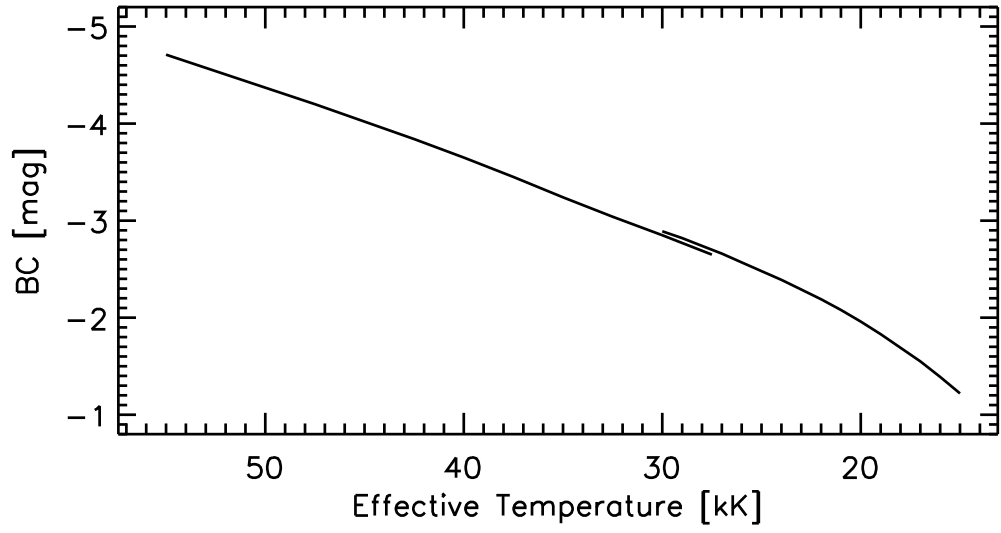}
\figurenum{18}
\caption{Bolometric corrections vs. effective temperature for solar composition,
          main-sequence OB stars. 
         \label{BCFig}}
\end{figure}

\clearpage

\begin{figure}
\includegraphics[scale=0.8,angle=0]{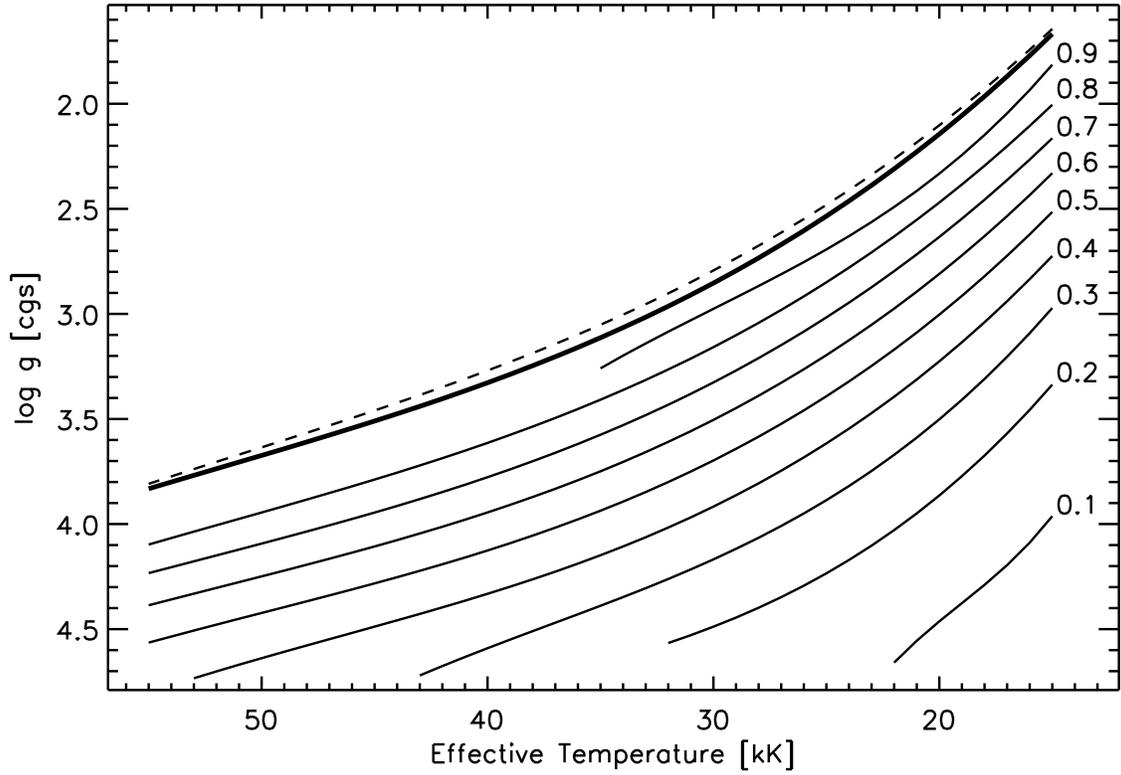}
\figurenum{19}
\caption{Isocurves of radiative acceleration relative to gravitational acceleration,
  $\Gamma_{\rm rad} = \max(g_{\rm rad}) / g$, for solar composition model atmospheres,
  as a function of \teff\  and $\log g$. $\Gamma_{\rm rad}$ values are listed right to
  the isolines. The thick line shows an estimate of the Eddington limit obtained by
  extrapolation; the dashed line corresponds to the estimate of the Eddington limit
  for metal-free model atmospheres. \label{EddingFig}}
\end{figure}

\end{document}